\documentclass[prd,twocolumn,nofootinbib,superscriptaddress,floatfix,preprintnumbers]{revtex4-2}
\usepackage{amsmath,amssymb,amsfonts,bm,graphicx,hyperref,xcolor,mathtools}
\usepackage[normalem]{ulem}
\usepackage{comment}
\usepackage{multirow}
\usepackage{xcolor}
\hypersetup{colorlinks=true, linkcolor=blue, citecolor=blue, urlcolor=blue}

\usepackage{mathtools,slashed,soul,xcolor}
\usepackage[english]{babel}

\newcommand{\lsim}{\mathrel{\mathop{\kern 0pt \rlap
  {\raise.2ex\hbox{$<$}}}
  \lower.9ex\hbox{\kern-.190em $\sim$}}}
\newcommand{\gsim}{\mathrel{\mathop{\kern 0pt \rlap
  {\raise.2ex\hbox{$>$}}}
  \lower.9ex\hbox{\kern-.190em $\sim$}}}

\newcommand{\sigmav}{\ensuremath{\langle\sigma v\rangle}}

\newcommand{\pp}{pp}
\newcommand{\pbar}{\ensuremath{\overline{\mathrm{p}}}}
\newcommand{\nbar}{\ensuremath{\overline{\mathrm{n}}}}
\newcommand{\dbar}{\ensuremath{\overline{\mathrm{D}}}}

\newcommand{\hebar}{\ensuremath{^3\overline{\mathrm{He}}}}
\newcommand{\hethreebar}{\ensuremath{^{3}\overline{\mathrm{He}}}}

\newcommand{\BIG}{{\sf BIG}}
\newcommand{\SLIM}{{\sf SLIM}}
\newcommand{\QUAINT}{{\sf QUAINT}}
\newcommand{\usine}{{\sc usine}}

\begin{document}

\title{Revisiting predictions for cosmic-ray antinucleon fluxes from Galactic Dark Matter}

\author{Lorenzo Stefanuto}
\email{lorenzo.stefanuto@unito.it}
\affiliation{Dipartimento di Fisica, Università di Torino, via P. Giuria, 1, 10125 Torino, Italy}
\affiliation{Istituto Nazionale di Fisica Nucleare, Sezione di Torino, Via P. Giuria 1, 10125 Torino, Italy}

\author{Mattia Di Mauro}\email{dimauro.mattia@gmail.com}
\affiliation{Istituto Nazionale di Fisica Nucleare, Sezione di Torino, Via P. Giuria 1, 10125 Torino, Italy}

\author{Fiorenza Donato}\email{fiorenza.donato@unito.it}
\affiliation{Dipartimento di Fisica, Università di Torino, via P. Giuria, 1, 10125 Torino, Italy}
\affiliation{Istituto Nazionale di Fisica Nucleare, Sezione di Torino, Via P. Giuria 1, 10125 Torino, Italy}
\affiliation{Theoretical Physics Department, CERN, 1211 Geneva 23, Switzerland}

\author{Nicolao Fornengo}\email{nicolao.fornengo@unito.it}
\affiliation{Dipartimento di Fisica, Università di Torino, via P. Giuria, 1, 10125 Torino, Italy}
\affiliation{Istituto Nazionale di Fisica Nucleare, Sezione di Torino, Via P. Giuria 1, 10125 Torino, Italy}

\author{Jordan Koechler}
\email{jordan.koechler@gmail.com}
\affiliation{Istituto Nazionale di Fisica Nucleare, Sezione di Torino, Via P. Giuria 1, 10125 Torino, Italy}

\author{David~Maurin}
\email{dmaurin@lpsc.in2p3.fr}
\affiliation{LPSC, Université Grenoble-Alpes, CNRS/IN2P3, 38026, Grenoble, France}

\begin{abstract}
The data on cosmic antiprotons have reached an outstanding precision on energies spanning from GeV to hundreds of TeV, thanks to the space-based \textsf{AMS-02} experiment. 
The balloon-borne \textsf{GAPS} experiment, which just completed its first Antarctic flight, will address antiproton and antideuteron fluxes well below GeV energies. 
Antinuclei in cosmic rays, as well as being produced by spallation reactions between cosmic-ray nuclei and the atoms of the interstellar medium, may hide contributions from exotic sources, such as particle dark matter annihilation in the Galaxy. 
In this paper, we present predictions for cosmic antiproton, antideuteron and antihelium fluxes both from secondary and dark matter origin. We use state-of-the-art production spectra, nuclear coalescence for antinuclei, and Galactic propagation models to derive upper limits on the dark matter annihilation cross-section from \textsf{AMS-02} antiproton data in different propagation scenarios (\textsf{BIG} and \textsf{QUAINT}). 
We quantify the impact of future \textsf{GAPS} data, showing that its sensitivity to sub-GV antiprotons could improve the $\langle\sigma v\rangle$ constraints by up to an order of magnitude for light DM ($m_{\chi} \lesssim 50$ GeV). For heavier antinuclei, the detection perspective with existing and upcoming experiments are derived for those scenarios consistent with \textsf{AMS-02} antiproton flux.
The detectability of such signals strongly depends on the experiment, the propagation model, and the hadronization tuning. Our analysis underscores the complementarity of antinuclei channels for indirect DM searches and the critical role of low-energy windows in constraining light DM candidates.
\end{abstract}

\maketitle

\section{Introduction}
The search for antimatter in space has reached the status of precision physics with the data on positrons and antiprotons 
collected firstly by the \textsf{PAMELA} satellite-borne experiment \cite{Adriani:2012paa,Adriani:2008zq,PAMELA:2008gwm,PAMELA:2017bna} and 
currently by \textsf{AMS-02} on board the International Space Station \cite{PhysRevLett.117.091103,AMS:2016oqu,AMS:2021nhj}. 
As for antinuclei, no detection has been established so far \cite{Abe:2012tz,PAMELA:2017bna,AMS:2021nhj,Osteria:2020lxn,vonDoetinchem:2020vbj,PhysRevLett.132.131001}. The strongest upper bound on the antideuteron flux has been fixed by the \textsf{BESS-Polar II} experiment \cite{PhysRevLett.132.131001} at $6.7\times10^{-5}\,\mathrm{(m^2\,s\,sr\,GeV}/n)^{-1}$ in the range of kinetic energies per nucleon $E_{k/n}\in[0.163,\,1.100]\,\mathrm{GeV}/n$. 
It is however expected that \textsf{AMS-02} could reach a sensitivity of $2$--$8\times10^{-7}\,\mathrm{(m^{2}\,s\,sr\,GeV}/n)^{-1}$ in the range $E_{k/n}\in[0.2,\,4]\,\mathrm{GeV}/n$ 
by about 2030 \cite{2008ICRC....4..765C, Oliva:JENAA:2024}.
The \textsf{GAPS} program \cite{Aramaki:2015laa,Osteria:2020lxn,vonDoetinchem:2020vbj} is designed to reach peak sensitivity comparable to \textsf{AMS-02} in the low-energy window $E_{k/n}\in[0.1,\,1]\,\mathrm{GeV}/n$ \cite{vonDoetinchem:2015yva,Aramaki:2015laa}. The balloon-borne \textsf{GAPS} exeriment has just completed its first successful flight (25 days) around Antarctica~\cite{gaps_Hailey_cern}, and two more flights are foreseen in the next years. 

The interest in cosmic-ray (CR) antinuclei is many-fold. 
Firstly, antiprotons are expected to be produced by spallation reactions of CR nuclei (mainly protons and helium) on the interstellar medium (ISM) atoms, the so-called secondary component. The detection of a flux of antiprotons from 1\,GeV  to beyond 400\,GeV probes  that antimatter is produced in the Galaxy by inelastic scatterings of CR nuclei off proton and helium in the ISM, at a rate compatible with measured hadronic cross-sections \cite{Donato:2008jk,Boudaud:2019efq,DiMauro:2023jgg,Calore:2022stf}. 
Secondly, CR light antinuclei, especially antideuterons (\dbar) \cite{Donato:1999gy} and antihelium (\hebar) \cite{Cirelli:2014qia,Carlson:2014ssa}, are considered among the cleanest messengers for indirect dark matter (DM) searches. The reason resides in the baryon-number conservation, which implies a strong 
suppression of the secondary flux at kinetic energy per nucleon $E_{k/n}\lsim 1~\mathrm{GeV}/n$
\cite{Chardonnet:1997dv,Donato:1999gy}. 
DM annihilation or decay in the Galactic halo, being non-baryonic particles, can annihilate into very low-energy antinuclei. 
Fluxes for the \dbar\ flux from standard Weakly Annihilating Massive Particles (WIMPs) are typically predicted two orders of magnitude higher than the secondary production expectations at low $E_{k/n}$ \cite{Donato:1999gy,Fornengo:2013osa,Korsmeier:2017xzj,Kachelriess:2020uoh,Serksnyte:2022onw,DeLaTorreLuque:2024htu}.

In this paper, we compute novel CR antiproton and antideuteron fluxes for both secondary production and DM annihilation, incorporating several updated theoretical and experimental ingredients. First, we model CR propagation with the \usine{} code~\cite{2020CoPhC.24706942M}, adopting state-of-the-art propagation setups derived in Refs.~\cite{Weinrich:2020cmw,Weinrich:2020ftb} from combined fits to several primary and secondary CR species. Second, we calculate the primary DM-induced \pbar{} fluxes using the source spectra provided by the \texttt{CosmiXs} package~\cite{Arina:2023eic,DiMauro:2024kml,Arina:2025ner}. These spectra are based on a dedicated tuning of \texttt{PYTHIA}~\cite{Bierlich:2022pfr} to a comprehensive set of LEP measurements. Third, we derive new upper limits on the DM annihilation cross section, \sigmav{}, by fitting the seven-year \textsf{AMS-02} antiproton dataset~\cite{AMS:2021nhj}. For the first time, we quantify the potential impact of simulated low-energy \pbar{} data from \textsf{GAPS} on these limits. Fourth, we implement novel differential production cross sections for secondary antideuterons \cite{DiMauro:2026oto}, and compute the source spectra for DM-induced antinuclei using the coalescence framework developed in Refs.~\cite{DiMauro:2024kml,DiMauro:2026oto}. Within this framework, the coalescence uncertainties are estimated to be approximately $20\%$, significantly smaller than those associated with previous modeling approaches. Combining these ingredients, we revisit the production and propagation of CR antinuclei in the Galaxy and assess the detection prospects for \dbar{} and \hebar{} in current and future experiments, including newly derived sensitivity projections for the proposed \textsf{ALADInO} experiment~\cite{instruments6020019}.

\section{Coalescence models}
\label{sec:coalmodels}

The production of light (anti)nuclei is commonly modeled via the \emph{coalescence} of (anti)nucleons: a bound state forms when the relevant (anti)nucleons are sufficiently close in phase space.
In phenomenological applications, this concept can be modeled in various ways, ranging from analytic prescriptions (useful for intuition and rapid estimates), to event-by-event afterburners interfaced with Monte Carlo (MC) generators, and ultimately to quantum-mechanical approaches based on Wigner functions. As most of the technical details and validation procedures are presented in Refs.~\cite{DiMauro:2024kml,DiMauro:2026oto}, we limit ourselves here to a concise summary of the implementations adopted in this work, highlighting the key assumptions and parameters.

\subsection{Analytic (spherical) coalescence}
\label{subsec:spherical}

The simplest approach assumes independent and isotropic emission of (anti)protons and (anti)neutrons, neglecting correlations in both momentum and configuration space \cite{Donato:1999gy,Korsmeier:2017xzj}. Coalescence is then encoded in a single momentum scale, the \emph{coalescence momentum} $p_{\rm coal}$. The invariant spectrum of a nucleus with mass number $A$ and $(Z,N)$ can be related to the corresponding single-(anti)nucleon spectra as
\begin{equation}
E_A \frac{d^3 N_A}{dp_A^3}
= B_A
\left( E_{\rm p} \frac{d^3 N_{\rm p}}{dp_{\rm p}^3} \right)^Z
\left( E_{\rm n} \frac{d^3 N_{\rm n}}{dp_{\rm n}^3} \right)^N ,
\end{equation}
with the momenta evaluated at $p_{\rm p,n}=p_A/A$. In practice one often sets $d^3N_{\rm n}/dp_{\rm n}^3 \simeq d^3N_{\rm p}/dp_{\rm p}^3$ (and similarly for antiparticles), so that for antideuterons one obtains
\begin{equation}
E_{\dbar} \frac{d^3 N_{\dbar}}{dp_{\dbar}^3}
= B_2 \left( E_{\pbar}\frac{d^3 N_{\pbar}}{dp_{\pbar}^3} \right)^2 .
\end{equation}
In the spherical approximation $B_A$ is proportional to $p_{\rm coal}^3$ \cite{Donato:1999gy,Korsmeier:2017xzj}. Experimentally, coalescence factors extracted from collider data are typically $B_2\sim \mathcal{O}(10^{-2})\,\mathrm{GeV}^2/c^3$ in $\pp$ collisions around $p_T/A\sim 1\,\mathrm{GeV}/c$, and often show an effective rise with $p_T/A$ in multiplicity-integrated samples \cite{ALICE:2020foi,ALICE:2025antideuteronRapidity13TeV}. Such trends can partly reflect genuine physics (e.g.\ changing emission volumes) and partly arise from spectral-shape effects when the proton and deuteron slopes are not self-similar \cite{ALICE:2019multiplicitypPbLightNuclei,ALICE:2020foi}. Because this framework ignores correlations, it is known to become unreliable precisely where correlations are strongest (e.g.\ at high transverse momentum or in boosted topologies \cite{Fornengo:2013osa,DiMauro:2024kml}). This motivates the use of event-by-event approaches.

\subsection{Monte-Carlo coalescence with phase-space cutoffs}
\label{subsec:cutoff_coal}

In MC afterburners, coalescence is applied directly to the event record and therefore automatically retains correlations generated by the underlying hadronization model \cite{Fornengo:2013osa,DiMauro:2024kml}. The most common prescription enforces a momentum-space proximity condition in the pair center-of-mass frame,
\begin{equation}
|\Delta\vec p|\equiv|\vec p_{\rm p}-\vec p_{\rm n}| < p_{\rm coal}\,,
\end{equation}
which we denote as the “$\Delta p$ model” \cite{Ibarra:2012cc,Fornengo:2013osa,Korsmeier:2017xzj,Kachelriess:2020uoh}. A minimal extension adds an explicit coordinate-space requirement \cite{Fornengo:2013osa},
\begin{equation}
|\Delta\vec r|\equiv|\vec r_{\rm p}-\vec r_{\rm n}| < r_{\rm coal}\,,
\end{equation}
with $r_{\rm coal}$ expected to be of the order of a few fm, comparable to the deuteron size \cite{CREMA:2016idx}. We refer to this model as the “$\Delta p+\Delta r$ model”. Besides capturing geometric information of the emitting source, the spatial cut is particularly useful to suppress coalescence between promptly produced antinucleons and antinucleons originating from displaced vertices (e.g., weak decays), since such pairs typically fail the $\Delta r$ condition \cite{DiMauro:2025vxp}. In both variants, the coalescence scale(s) are treated as effective parameters to be calibrated on collider measurements.

\subsection{Wigner-function coalescence}
\label{subsec:wigner_model}

A more microscopic formulation embeds coalescence into quantum mechanics by expressing the bound-state formation probability as an overlap in phase space between the two-particle density matrix of the \pbar{}--\nbar{} system and the density matrix of the deuteron bound state \cite{Scheibl:1998tk,Bellini:2018epz,Blum:2017qnn,Bellini:2020cbj,Kachelriess:2019taq,Mahlein:2023fmx}. In Wigner language, the differential yield can be written schematically as
\begin{equation}
\frac{d^3 N_{\rm D}}{dp_{\rm D}^3}
=
S \int \frac{d^3 r\, d^3 q}{(2\pi)^3}\;
\mathcal{D}(\vec r,\vec q)\;
W_{\rm pn}\!\left(\tfrac{\vec p_{\rm D}}{2}+\vec q,\tfrac{\vec p_{\rm D}}{2}-\vec q;\,\vec r\right),
\label{eq:wigner_short}
\end{equation}
where $S$ accounts for spin/isospin statistics (for \dbar{}, $S=3/8$), $\vec r\equiv \vec r_{\rm p}-\vec r_{\rm n}$ is the relative separation, and $\vec q\equiv (\vec p_{\rm p}-\vec p_{\rm n})/2$ so that $\Delta\vec p = 2\vec q$.
The kernel $\mathcal{D}(\vec r,\vec q)$ is the deuteron Wigner function, fully determined by the bound-state wavefunction, while $W_{\rm pn}$ is the pair Wigner function encoding the correlated two-particle phase-space distribution provided by the event generator.

In practice, we estimate the momentum correlations directly from MC events and approximate the spatial part with a source model extracted from the same generator, following the strategy of Ref.~\cite{DiMauro:2024kml}. A convenient working assumption is to factorize
\begin{equation}
W_{\rm pn} \simeq H_{\rm pn}(\vec r_{\rm p},\vec r_{\rm n})\,G_{\rm pn}(\vec p_{\rm p},\vec p_{\rm n}),
\end{equation}
where $G_{\rm pn}$ is obtained from the two-particle momentum distribution in the event record (including correlations), and $H_{\rm pn}$ encodes the effective space--time distribution of emission points. While this neglects genuine two-particle spatial correlations beyond those induced by the common event geometry, it makes the model transparent and allows controlled variations (e.g.\ different vertex prescriptions) to probe the corresponding uncertainty.

For an event-by-event implementation, Eq.~\eqref{eq:wigner_short} can be interpreted as defining an acceptance weight in $(\Delta\vec r,\Delta\vec p)$ space. Concretely, for each \pbar{}--\nbar{} pair in an event we evaluate the kernel value $\mathcal{D}(\Delta r,\Delta p)$ (and the source factor, when needed), draw a random number $u\in[0,1]$, and accept the pair to form an antideuteron if $u < \mathcal{W}(\Delta r,\Delta p)$, where $\mathcal{W}$ is a suitably normalized weight. This is the same sampling logic adopted in Refs.~\cite{DiMauro:2024kml,DiMauro:2026oto}.

\paragraph{Gaussian kernel.}
As a compact benchmark we consider a Gaussian bound-state wavefunction, for which the Wigner kernel is analytic \cite{Kachelriess:2020uoh}. In this case the phase-space overlap effectively selects pairs with $\Delta r \lesssim \mathcal{O}({\rm few~fm})$ and $\Delta p \lesssim \mathcal{O}(0.1\text{--}0.2\,{\rm GeV})$, with the precise width controlled by a single momentum/length scale. 
The Gaussian implementation is flexible and numerically efficient, but it typically exhibits a degeneracy between the assumed source size and the bound-state scale; in practice this is broken by fixing the source size to a benchmark value (or by importing it from the generator), while tuning the remaining parameter to collider data, as done in Ref.~\cite{DiMauro:2024kml} and in the present work.

\paragraph{Argonne kernel.}
To reduce the arbitrariness associated with purely phenomenological kernels, we also consider a Wigner kernel constructed from a realistic deuteron wavefunction based on the Argonne $v_{18}$ nucleon--nucleon potential \cite{Wiringa:1994wb}. This wavefunction is constrained by scattering data and by the deuteron binding energy, and therefore provides a physically motivated reference without introducing an ad-hoc coalescence momentum. The corresponding Wigner kernel does not admit a simple closed form; we employ a tabulated representation as a function of $(r,q)$ based on the parametrization in Ref.~\cite{Horst:2023oti}. Once the underlying antinucleon production is fixed (through the MC tune), the Argonne/Wigner approach is comparatively predictive and serves as a useful cross-check of the calibrated phenomenological models \cite{DiMauro:2024kml}.

Overall, these three classes of prescriptions (analytic, cutoff afterburners, and Wigner-based coalescence) allow us to: (i) connect to the standard literature, (ii) retain generator-level correlations relevant for collider and CR applications, and (iii) quantify the residual model dependence associated with the bound-state kernel and the effective source description.

\section{Production of primary antinuclei from dark matter annihilation} 
\label{sec:prim_production}

Primary antinuclei can be produced through DM annihilation in the Galactic halo. The corresponding production rate of an antinucleus species $i$, per unit volume, time, and kinetic energy per nucleon $E_{k/n,i}$ (the so-called source term) can be written---assuming annihilation into a single channel---as:
\begin{equation}
\label{eq:primsource}
    q_i^{\rm prim}(E_{k/n,i},\vec{x}) = \frac{\sigmav}{2\xi m_\chi^2}\rho_\chi^2(\vec{x}) \frac{dN_i}{dE_{k/n,i}}\;,
\end{equation}
where \sigmav{} is the thermally averaged annihilation cross-section, $m_{\chi}$ is the DM particle mass, and $\xi =1$ if DM particles are self-conjugate and 2 otherwise. The Galactic DM density distribution is described by $\rho_\chi$. Following Refs.~\cite{Calore:2022stf,DeRomeri:2025dwm}, we adopt a spherically symmetric Navarro-Frenk-White (NFW) profile~\cite{Navarro:1995iw}, with a scale radius $r_s=19.6$~kpc~\cite{McMillan_2016}, local DM density $\rho_\odot = 0.385$\,GeV/cm$^3$~\cite{2021RPPh...84j4901D}, and solar distance $R_\odot = 8.20$~kpc~\cite{2019AandA...625L..10G}. Finally, $dN_i/dE_{k/n,i}$ denotes the differential yield of antinuclei at production for the specified DM annihilation channel. 

To reliably predict the antinucleon differential yield, we employ a dedicated tuning of \texttt{PYTHIA}~\cite{Bierlich:2022pfr}, an MC event generator for high-energy particle collisions. This tuning specifies hadronization parameters and branching ratios and is calibrated to reproduce key observables measured in $e^+e^-$ collisions at the $Z$-pole, which provide a clean environment that closely resembles the hadronization conditions relevant for DM annihilation. The differential yields of \dbar{} and \hebar{} are then obtained by applying the coalescence criteria described in Sect.~\ref{sec:coalmodels} as an external afterburner to the \texttt{PYTHIA} events. Specifically, in order to form \hebar, the criteria must be valid for each of the three antinucleon pairs in the $\pbar\,\pbar\,\nbar$ configuration.

As an example, \texttt{CosmiXs}~\cite{Arina:2023eic,DiMauro:2024kml,Arina:2025ner} provides predictions for \pbar{} and $\overline{\mathrm{D}}$ spectra from DM annihilation, relying on a calibrated \texttt{PYTHIA} setup that reproduces the $\overline{\mathrm{D}}$ multiplicity measured by the \textsf{ALEPH}~\cite{ALEPH:2006qoi} collaboration. It also accurately describes a range of meson and baryon observables from \textsf{ALEPH} and other $e^+e^-$ experiments such as \textsf{DELPHI}, \textsf{L3}, and \textsf{OPAL} (see Refs.~\cite{Jueid:2022qjg,Jueid:2023vrb,Amoroso:2018qga} for further details).

More recently, Ref.~\cite{DiMauro:2025vxp} extended this calibration to better predict the \dbar{} and \hebar{} differential yields. In particular, the tuning was refined to reproduce the fragmentation fraction of $b$ quarks into weakly decaying $b$-baryons, $f(b\to \Lambda_b)=0.089\pm0.012$, as reported by the \textsf{HFLAV} collaboration~\cite{HFLAV:2019otj}, as well as the recent upper limit on the inclusive branching ratio $\mathrm{BR}(\overline{\Lambda^0_b}\to{}^3\overline{\mathrm{He}}X)<6.3\times10^{-8}$ set by the \textsf{LHCb} experiment~\cite{Moise:2024wqy}. Accurately describing these observables is important, as the decay of weakly decaying $B$-hadrons constitutes a significant source of antinucleons, and hence of antinuclei. In the following, we refer to this setup as the “$\Lambda_b$ tuning”.

For primary antinuclei, the choice of coalescence model has a negligible impact on the predicted $\overline{\mathrm{D}}$ differential yield, $dN_{\overline{\mathrm{D}}}/dE_{k/n,\overline{\mathrm{D}}}$. As shown in Ref.~\cite{DiMauro:2024kml}, all considered coalescence models provide a good description of the $\overline{\mathrm{D}}$ multiplicity measured in $e^+e^-$ collisions at the $Z$-pole by the \textsf{ALEPH} collaboration. The main limitation arises from the relatively large experimental uncertainty of this measurement, at the level of $\sim30\%$. As a benchmark, we therefore adopt the Argonne Wigner model, which is among the most physically motivated approaches discussed in Sect.~\ref{sec:coalmodels}.

\section{Production of secondary antinuclei}
\label{sec:sec_production}

Secondary antinuclei are produced through inelastic collisions between primary CRs and nuclei in the ISM. In principle, all CR species—from protons (p) and helium (He) up to heavier nuclei such as iron (Fe)—can contribute to secondary antinuclei production. However, CR species with charge $Z>2$ are significantly less abundant, by at least two (one) orders of magnitude compared to protons (helium). Furthermore, the ISM is predominantly composed of hydrogen and helium. As a result, the dominant production channels for secondary antinuclei involve CR protons and helium nuclei colliding with hydrogen and helium targets.

The source term for secondary antinuclei of species $i$ can be written as
\begin{multline}
    \label{eq:secsource}
    q_i^{\rm sec}(E_{k/n,i},\vec{x})=\sum_{\rm c,t}\int_{E_{k/n,\rm c}^{\rm min}}^\infty dE_{k/n,\rm c}\;4\pi\,n_{\rm t}(\vec{x})\times \\ 
    \times \Phi_{\rm c}(E_{k/n,\rm c},\vec{x}) \frac{d\sigma_{{\rm ct}\to i}}{dE_{k/n,i}}(E_{k/n,\rm c})\;,
\end{multline}
where $n_{\rm t}$ denotes the number density of ISM target nuclei, $\Phi_{\rm c}=v\,n_{\rm c}$ is the interstellar flux of CR primaries, and $d\sigma_{{\rm ct}\to i}/dE_{k/n,i}$ is the differential production cross-section for the secondary antinucleus $i$. 

In Eq.~\eqref{eq:secsource}, the lower integration limit $E_{k/n,\rm c}^{\rm min}$ represents the minimum kinetic energy per nucleon of the primary CR required to produce an antinucleus at rest, as dictated by baryon number conservation. Table~\ref{tab:thresholdK} summarizes the corresponding threshold energies for different production channels and antinuclei species.

\begin{table}[]
    \centering
    \caption{Threshold kinetic energy per $n$ of the primary CR projectile ($\rm p$, He) to produce secondary antinuclei (\pbar{}, \dbar{}, $^{3}\overline{\rm He}$) by colliding with nuclei from the ISM (H and He).}
    \label{tab:thresholdK}
    \begin{tabular}{|c|c|c|c|}
        \cline{2-4}
        \multicolumn{1}{c|}{} & \multicolumn{3}{|c|}{$E_{k/n, \rm c}^{\rm min}$ [GeV/n]} \\
        \hline
        Process & secondary \pbar{} & secondary \dbar{} & secondary ${}^3\overline{\rm He}$ \\
        \hline
        p+H & $6\,m_{\rm p}$ & $16\,m_{\rm p}$ & $30\,m_{\rm p}$ \\
        \hline
        p+He & \multirow{2}{*}{$3\,m_{\rm p}$} & \multirow{2}{*}{$7\,m_{\rm p}$} & \multirow{2}{*}{$12\,m_{\rm p}$} \\
        He+H & & & \\
        \hline
        He+He & $1.125\,m_{\rm p}$ & $2.5\,m_{\rm p}$ & $4.125\,m_{\rm p}$ \\
        \hline
    \end{tabular}
\end{table}

In Ref.~\cite{DiMauro:2026oto}, the differential production cross-section for antideuterons, $d\sigma_{\mathrm{pH}\to\dbar}/dE_{k/n,\,\dbar}$, is derived using a dedicated tuning of \texttt{PYTHIA}, optimized to reproduce \pbar{} measurements in pp collisions. When combined with the coalescence models described in Sect.~\ref{sec:coalmodels}, this tuning provides a good description of \dbar{} and \hebar{} production as measured by the \textsf{ALICE} experiment~\cite{ALICE:2017xrp, ALICE:2020foi}. Within this framework, the differential cross-section can be written as
\begin{equation}
\frac{d\sigma_{\rm pH\to \dbar}}{dE_{k/n,\,\dbar}}(E_{k/n,\rm p})
=\sigma_{\rm pp}^{\rm inel}(E_{k/n,\rm p})\,
\frac{dN_{\dbar}}{dE_{k/n,\, \dbar}}(E_{k/n,\rm p})\;,
\end{equation}
where $\sigma_{\rm pp}^{\rm inel}$ is the total inelastic proton–proton cross-section (for which we adopt the parametrization of Ref.~\cite{Orusa:2022pvp}), and $dN_{\dbar}/dE_{k/n,\dbar}$ denotes the antideuteron spectrum produced per collision, as obtained from the MC simulations.

This formulation can be generalized to arbitrary CR projectiles and ISM targets by parameterizing the total inelastic cross-section through an empirical mass-scaling relation~\cite{Orusa:2022pvp}
\begin{equation}
\sigma_{\rm ct}^{\rm inel}(E_{k/n,\rm c}) \simeq (A_{\rm c}A_{\rm t})^{0.8}\,\sigma_{pp}^{\rm inel}(E_{k/n,\rm c})\;,
\end{equation}
where $A_{\rm c}$ and $A_{\rm t}$ are the mass numbers of the CR projectile and the ISM target, respectively.

In this work, we compute the source term for secondary \pbar{} in Eq.~(\ref{eq:secsource}) according to the results in \cite{Korsmeier:2018gcy} and already implemented in \cite{Boudaud:2019efq,Calore:2022stf}. The source term for secondary $\dbar$ is computed as in Ref.~\cite{DiMauro:2026oto}, whose cross-sections are publicly available in the Zenodo repository~\cite{di_mauro_2026_19099608}. As a benchmark, we adopt the results from the Gaussian Wigner coalescence model. In addition to being physically well motivated, this model provides the second-best fit to the \textsf{ALICE} $\overline{\mathrm{D}}$ data~\cite{DiMauro:2026oto}. The best fit is achieved by the simple $\Delta p$ model; however, as discussed in Ref.~\cite{DiMauro:2026oto}, this model is disfavored for applications to astrophysical $\overline{\mathrm{D}}$ production, since it may yield a non-negligible fraction of non-physical $\overline{\mathrm{D}}$.

\section{Galactic propagation and solar modulation}
\label{sec:propagation}

\subsection{Transport equation and source terms}
We consider the CRs diffusing in a cylindrical volume of radius $R_G=20$\,kpc and half-height $L$. The astrophysical sources and gas\footnote{We take the ISM density to be $n_{\rm ISM}=1~$cm$^{-3}$ with 90\% H and 10\% He in number, and $\langle n_e\rangle=0.033$~cm$^{-3}$ and $T_e=10^4$~K.} are located in a thin disc of half-height $h=100$\,pc, where nuclear interactions and continuous energy losses and gains occur. The steady-state transport equation for the differential density $n$ of Galactic CR antinuclei \cite{1990acr..book.....B,2002cra..book.....S,Strong:2007nh} in the thin disc approximation, assuming a constant velocity $V_c$ perpendicular to the Galactic plane and isotropic and space-independent diffusion coefficient $K(E)$ as a function of total energy, reads \cite{Maurin:2018rmm}:
\begin{eqnarray}
    &-&\left[K\Delta - V_{c} \frac{\partial}{\partial z}\right] n 
    +2h\,\delta(z) \frac{\partial}{\partial E} \left[ b(E)\,n  - c(E)\, \frac{\partial n}{\partial E} \right]
     \nonumber\\
     &=& q^{\rm prim} \!+ 2h\delta(z) \Big[q^{\rm sec}\!+q^{\rm ter}
     -\sum_{t\in \rm ISM}\!\!\! n_t v\,\sigma_{\rm inel}\;n\Big]\,.
     \label{eq:transport}
\end{eqnarray}
In the first line, $\Delta$ is the Laplacian operator in cylindrical coordinates and $b(E)$ is the first-order energy redistribution term,
\begin{equation}
  b(E) = \left\langle\frac{dE}{dt}\right\rangle-E_k\left(\frac{2m+E_k}{m+E_k}\right) \frac{V_c}{3h}+(1+\beta^{2}) \frac{K_{pp}}{E},
\end{equation}
which includes ionization losses on neutral hydrogen and Coulomb losses in the ionized ISM (first term), adiabatic losses (second term) and reacceleration (last term). There is also $c(E)=\beta^{2} K_{pp}$, the second-order term for reacceleration, which depends on the momentum diffusion term,
\begin{equation}
K_{pp} =\frac{4}{3}\frac{V_a^2\beta^2 E^2}{K(R)} \frac{1}{\delta(4\!-\!\delta^2)(4\!-\!\delta)}\,,
 \end{equation} 
mediated by an effective Alfvén velocity $V_a$ \cite{1994ApJ...431..705S}, and with $\delta$ the slope of the diffusion coefficient $K(R)$ (see below) in the low energy regime.

In addition to the primary and secondary source terms, $q^{\rm prim}$ and $q^{\rm sec}$, defined respectively in Eqs.~\eqref{eq:primsource} and~\eqref{eq:secsource}, the transport Eq.~\eqref{eq:transport} also includes the tertiary source term
\begin{multline}
    \label{eq:tersource}
    q^{\rm ter}_i(E_i,\vec{x})=\sum_{\rm t}\Biggl[\int_{E_i}^\infty dE_i'\frac{d\sigma^{\rm ina}_{i{\rm t}\to i}}{dE_i}(E_i')\,v_i'\,n_i(E_i')- \\
    -\sigma^{\rm ina}_{i{\rm t}\to i}(E_i)\,v_i\,n_i(E_i)\Biggr]n_{\rm t}(\vec{x})\;,
\end{multline}
which accounts for the redistribution in energy of antinuclei (both primary and secondary). The first term describes the scattering of antinuclei species $i$, from high energies $E_i'$ down to energy $E_i$, through non-annihilating interactions, characterized by the differential inelastic non-annihilating cross-section $d\sigma^{\rm ina}_{i{\rm t}\to i}/dE_i$. The second term describes the scattering of antinuclei, from energy $E_i$ to lower energies, and is determined by the total inelastic non-annihilating cross-section $\sigma^{\rm ina}_{i{\rm t}\to i}$.

\subsection{IS flux calculation: transport parameters (and configurations)}
\label{subsect: IS flux calculation}

The transport Eq.~\eqref{eq:transport} is solved semi-analytically with the \usine{} code\footnote{\url{https://dmaurin.gitlab.io/USINE/}} \cite{Maurin:2018rmm}. The differential density per kinetic energy per nucleon, calculated at Earth position, $n_\odot=n(z=0,r=R_\odot)$, is converted into an interstellar (IS) flux via (assuming isotropy), $\Phi_{\rm IS} = v/(4\pi) \times n_\odot$.

The IS flux depends on the transport parameters $K(R)$, $V_a$, $V_c$ and $L$: the combination of parameters $(K/L\,, V_a\,, V_c$) are determined from secondary-to-primary ratios \cite{Maurin:2001sj} (e.g., B/C), and the degeneracy between the normalisation of $K$ and $L$ is lifted from the use of CR clocks \cite{Donato:2001eq} (e.g., $^{10}$Be/$^9$Be). 
Following Ref.~\cite{2019PhRvD..99l3028G}, the diffusion coefficient is taken to be a generic power law with a possible low-energy break (or alternatively a modification in the non-relativistic regime) and another one at high-rigidity:
\begin{eqnarray}
\label{eq:def_K}
K(R) &=& K_0 \times \underbrace{\beta^\eta \;\times\; \left\{ 1 + \left( \frac{R}{R_{\rm l}} \right)^{\frac{\delta_{\rm l}-\delta}{s_{\rm l}}}
  \right\}^{s_{\rm l}}}_{\text{non-relativistic $\times$ low-rigidity break}}\,\\
&\times& \underbrace{\!\Biggl\{  \frac{R}{\left(R_0 = 1\,{\rm GV}\right)} \Biggl\}^\delta\!\!}_{\text{intermediate}} \,
\times \,\,\underbrace{\!\left\{  1 + \left( \frac{R}{R_{\rm h}} \right)^{\frac{\delta-\delta_{\rm h}}{s_{\rm h}}}
  \right\}^{\!\!-s_{\rm h}}\!\!\!\!}
_{\text{high-rigidity break}}\,.\nonumber
\end{eqnarray}
In Ref.~\cite{2019PhRvD..99l3028G}, the degeneracies when fitting this large number of free parameters was mitigated by 
adjusting separately the high-rigidity break parameters ($R_h,\,\delta_h,\,s_h$)---whose uncertainties were found not to impact the lower-energy parameters---, by fixing the unconstrained smoothness of the low-rigidity break parameter to $s_l=0.04$ (fast transition), and by exploring three benchmark configurations with different numbers of free parameters. Ref.~\cite{2019PhRvD..99l3028G} found that the current CR data were fit equally well by these configurations. Two of them have 6 free parameters, namely \SLIM{} ($K_0$, $L$, $\delta$, $R_l$, $\delta_l$, $s_l$, enforcing $V_a=V_c=0$) and \QUAINT ($K_0$, $L$, $\delta$, $\eta$, and $V_c$, $V_a$, enforcing no low-rigidity break), whereas \BIG{} has 8 parameters ($K_0$, $L$, $\delta$, $R_l$, $\delta_l$, $s_l$, $V_c$, $V_a$, enforcing $\eta=1$).

The value of all the above parameters (and uncertainties) are taken from a fit to (Li,Be,B)/C and $^{10}$Be/$^9$Be data \cite{Weinrich:2020cmw,Weinrich:2020ftb}. 
The secondary antiproton flux is calculated from the procedure detailed in Ref.~\cite{Boudaud:2019efq}: the Galactic transport parameters are obtained from the fit to the above-mentioned secondary-to-primary ratios, and the primary source parameters (injection normalizations and slopes) from the fit to the H, He, C, and O primary fluxes measured by AMS-02. To rigorously estimate the theoretical uncertainties on the antiproton background, $\approx 10^4$ parameter realizations were drawn from the covariance matrices of the transport parameters, primary nucleus fluxes, and antiproton production cross-sections.
The main advantage of this approach is that the dispersion of this ensemble keeps track of the full covariance of systematics uncertainties on the modelled background. 
The latter is critical to have a statistically sound and relevant assessment of the presence or absence of a DM primary component in the data \cite{Boudaud:2019efq,Calore:2022stf}. As heavier CR antinuclei have not been detected yet, we do not repeat the time-consuming steps to have their associated covariance matrices, but consider the total uncertainties only. To calculate them, a first good approximation is to add quadratically the specific production cross-section uncertainties of the antinuclei (see Sect.~\ref{sec:sec_production}) to the transport and (propagated) primary flux uncertainties of the secondary antiprotons (as shown in Fig.~2 of Ref.~\cite{Boudaud:2019efq}).

For the calculation of the primary component, we follow closely the steps described in \cite{Calore:2022stf}. Namely, we use the source term discussed in Sect.~\ref{sec:prim_production} to calculate the fluxes on a grid of relevant DM parameters and $L$ values (the latter being only loosely constrained, while critical in the calculation \cite{Donato:2003xg,Genolini:2021doh}). We then rely on interpolations of these tabulated fluxes, so that we can explore different configurations and comparisons to the data in at most a minute (see next section).

\begin{figure}
    \centering
    \includegraphics[width=1\linewidth]{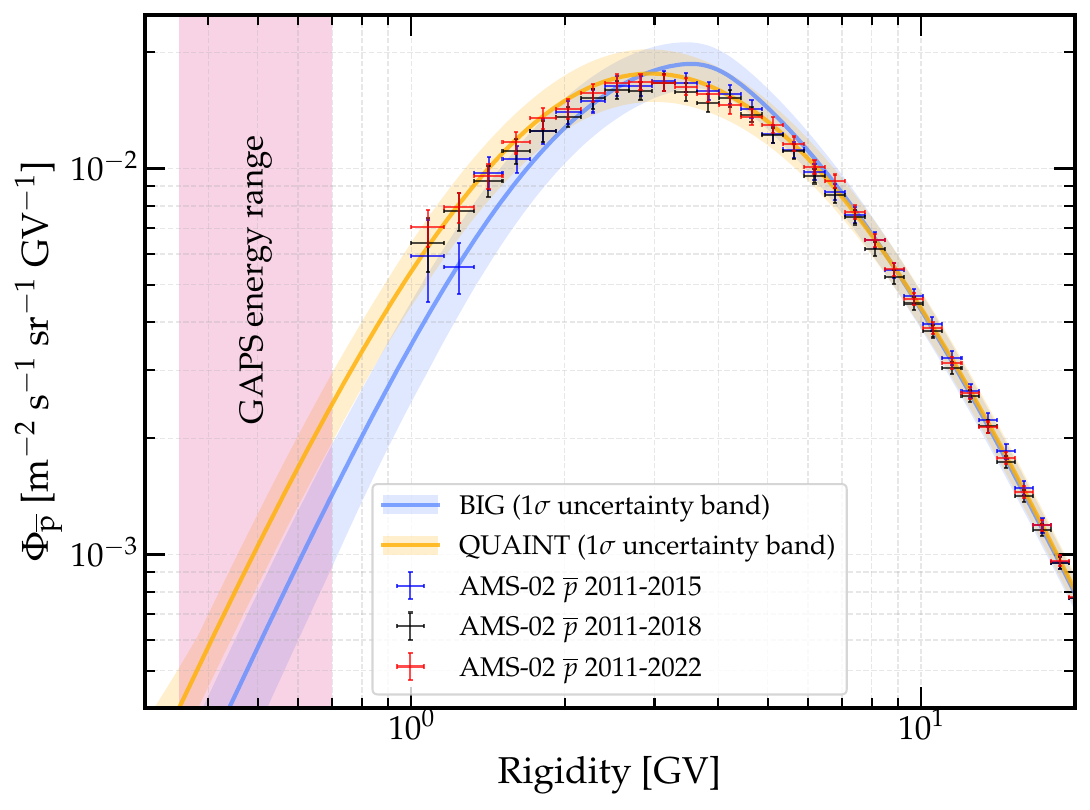}
    \caption{\textbf{TOA secondary antiproton fluxes} computed with \BIG{} and \QUAINT{} propagation models, alongside their $1\sigma$ theoretical uncertainty bands. Solar modulation is modeled considering $\phi_{\rm FF} = 678$\,MV, appropriate for the 2011-2018 period (see text). Data points are from \textsf{AMS-02}, for three different data taking periods: 2011--2015 (blue symbols) \cite{AMS:2016oqu}, 2011--2018 (black symbols) \cite{AMS:2021nhj}, and 2011--2022 (red symbols) \cite{PhysRevLett.134.051002}. The magenta vertical band shows the \textsf{GAPS} energy range for \pbar{} detection.}
    \label{fig:secondary_pbar_flux}
\end{figure}

There are several improvements in our analysis, compared to that of \cite{Calore:2022stf}, beside the extension of the calculation to heavier antinuclei:
\begin{itemize}
  \item First, we updated the DM source term (Sect.~\ref{sec:prim_production}), which however has a minor impact on the primary antiproton flux prediction.
  \item Second, compared to the \usine{} version used in Ref.~\cite{Calore:2022stf} to  constrain DM from antiprotons, we fixed a bug for the primary component calculation, slightly impacting the limits derived at low energy; we stress that this bug was already fixed in a recent analysis to set constraints on PBH-induced antiprotons and antideuterons~\cite{DeRomeri:2025dwm}.
  We also include in \usine{} the propagation of \hethreebar{}, whose inelastic cross-section on H and He is taken from Ref.~\cite{ALICE:2022zuz}.
  \item Third, the default transport configuration used for the secondary and primary calculations was taken to be \BIG{}. While this configuration gave a perfect match (statistically-wise) between the secondary production and the \textsf{AMS-02} antiproton first data release (four years of data, \cite{AMS:2016oqu}), it was already shown in Ref.~\cite{Calore:2022stf} to slightly overshoot (visually) and provide a less satisfactory fit to the latest complete \textsf{AMS-02} dataset (seven years of data, \cite{AMS:2021nhj}). We find here (see Fig.~\ref{fig:secondary_pbar_flux} and next section) that the secondary flux in the \QUAINT{} configuration gives a fairer visual match to these data\footnote{For the danger of drawing hasty conclusions based on the visual inspection that do not account for the covariance matrix of uncertainties (and for alternative safer visual representations), we refer the reader to \cite{Boudaud:2019efq} and their Fig.~2.}, and it also provides a slightly better fit to the data. Looking more closely at the $\chi^2_{\rm min}$ of these configurations, we recall that they were performing similarly on the secondary-to-primary (Table~1 of \cite{Weinrich:2020cmw}) and radioactive (Table~3 of \cite{Weinrich:2020ftb}) species. Nonetheless, the main visual difference between \QUAINT{} and the two other configurations is that the former displays a smoother transition than the latter, as reflected from the fact that $K(R)$ has no break (see Fig.~3 of \cite{Weinrich:2020cmw}).
\end{itemize}

\subsection{TOA fluxes: Solar modulation}

To obtain the TOA fluxes, we modulate the IS fluxes using the force-field approximation (FFA) \cite{GleesonEtAl1968a,Fisk1971}. 
The Fisk parameter corresponding to the \textsf{AMS-02} antiproton data-taking periods is calculated from neutron monitor data~\cite{Maurin:2014bva, GhelfiEtAl2017} and retrieved from the Cosmic-Ray Data Base (CRDB) interface\footnote{\url{https://lpsc.in2p3.fr/crdb/}}~\cite{MaurinEtAl2014, MaurinEtAl2020, MaurinEtAl2023}. The first \textsf{AMS-02} antiproton dataset was taken between 2011/05/19 and 2015/05/26~\cite{AMS:2016oqu}, while the second covers the period from 2011/05/19 to 2018/05/19~\cite{AMS:2021nhj}. A third dataset has been published over the 11-year solar cycle 2011--2022~\cite{PhysRevLett.134.051002}, but since it is kinematically limited up to $\approx 40$\,GV, we maintain the 2011--2018 release as our reference dataset. Focusing on the two broader datasets, the measurements from the \textsf{KIEL}, \textsf{OULU}, \textsf{ROME}, and \textsf{SOPO} stations yield an average Fisk potential of $\phi_{\rm FF} \approx 722 \pm 25$~MV for the 2011--2015 period, and $\phi_{\rm FF} \approx 678 \pm 25$\,MV for the 2011--2018 period; hence a difference of 44\,MV between the two datasets. This is smaller than the 100\,MV uncertainty from the NM analysis \cite{GhelfiEtAl2017}. We recall that the latter uncertainty was included as a nuisance parameter in the determination of the transport parameters \cite{Weinrich:2020cmw}. As a result, this uncertainty is automatically accounted for in the covariance matrix of the model uncertainties.

A more critical issue is that of the validity of using, as we do, the same modulation potential for differently charged particles. The Bartels-rotation averaged \textsf{AMS-02} proton, antiproton, electron, and positron data over 11 years (corresponding to 139 Bartels rotations) \cite{PhysRevLett.134.051002} enable to test the limitation of the FFA \cite{John:2025jln} or the successfulness of its simplest extensions \cite{Kuhlen:2019hqb,Cholis:2020tpi,Long:2024nty} and full modelling (e.g., Refs.~\cite{Aslam:2023gjv,Tomassetti:2025nna}). Ref.~\cite{John:2025jln} highlighted that up to November 2015, the Fisk potential to fit proton and antiprotons differ by $\lesssim 80$\,MV, but by more than 200\,MV after. Over the 7 years of the most precise antiproton dataset, this amounts to a $\sim 110$\,MV difference. This is within the above-mentioned uncertainty band, which, we recall, is smaller than the rest of the uncertainties. However, this is starting to reach an uncomfortable difference, and future \textsf{AMS-02} release including data from periods of high-Solar activity will render this issue more acute, not to say critical.

It is interesting to stress that the analyses in Refs.~\cite{John:2025jln,Duan:2025ead} both point at the importance of using at least the simplest extensions of the FFA, owing to its role in the significance of a potential DM signal in \textsf{AMS-02} data. However, the analysis of Ref.~\cite{Duan:2025ead} indicates that accounting for energy correlations in the data mitigates the impact of using different Solar modulation models (including the FFA). At this stage, the comprehensiveness of our analysis, which accounts for the propagation of cross-section, propagation and even Solar modulation uncertainties (though in the limited FFA), and also for the covariance matrix of systematic uncertainties on the data (in a more data-driven way than in \cite{Duan:2025ead}), does not call for an improved treatment of (the still debated) Solar modulation effects. \\

Fig.~\ref{fig:secondary_pbar_flux} displays the solar-modulated fluxes of secondary antiprotons predicted with the \textsf{BIG} and \textsf{QUAINT} propagation models. The $1\sigma$ theoretical uncertainties, discussed in Sect.~\ref{subsect: IS flux calculation}, are shown as colored bands. For comparison, we overlay all the available \textsf{AMS-02} datasets from different data-taking periods~\cite{AMS:2016oqu, AMS:2021nhj,PhysRevLett.134.051002}. The solar modulation is modeled considering $\phi_{\rm FF} = 678$\,MV, to maintain consistency with the 2011--2018 \textsf{AMS-02} dataset~\cite{AMS:2021nhj}. Finally, the shaded vertical band highlights the \pbar{} rigidity window targeted by the \textsf{GAPS} experiment.

\section{Results on antiproton fluxes}
\label{sec:results_antip}

Relying on the propagation framework, production cross-sections, and primary source spectra established in the previous sections, we now have all the ingredients to evaluate the primary and secondary TOA fluxes for antiprotons and, more generally, heavier antinuclei. Focusing first on the antiproton channel, we start by deriving updated upper limits on the DM annihilation cross-section, \sigmav{}, as detailed below.

\subsection{Derivation of upper limits on \sigmav\ from antiproton data: methodology}
\label{subsec:sigmav_UL}
To compute updated $95\%$ confidence level (CL) upper limits on \sigmav{}, we follow the statistical procedure outlined in Ref.~\cite{Calore:2022stf} (see also Ref.~\cite{DeRomeri:2025dwm} for a similar application), considering the \textsf{AMS-02} 2011--2018 antiproton dataset~\cite{AMS:2021nhj}. We briefly recall the main steps below.

The primary antiproton fluxes depend on $m_\chi$, \sigmav{}, and the propagation parameters, most notably the diffusion halo half-height $L$~\cite{G_nolini_2021}. Because the primary signal scales linearly with \sigmav{}, we optimize the analysis by precomputing a dense grid of flux predictions spanning the relevant parameter space of $m_\chi$ (5\,GeV--10\,TeV) and $L$ (1--12\,kpc).
We then define a log-likelihood function that combines the $\chi^2$ statistic with a penalty term for $L$:
$$ -2 \ln \mathcal{L}(L, \mu) = \sum_{i,j} x_i (\mathcal{C}^{-1})_{ij} x_j + \left( \frac{\log L - \log \hat{L}}{\sigma_{\log L}} \right)^2 \, , $$
where $x_i = \Phi_i^{\rm exp} - \Phi_i^{\rm th}(L, \mu)$, with $\Phi_i$ denoting the flux value in the $i$-th energy bin. The parameter array $\mu$ collectively represents the DM properties under investigation, namely \sigmav{} and $m_\chi$. The covariance matrix $\mathcal{C}$ accounts for both data uncertainties (statistical and systematic) and theoretical uncertainties on the modeled background (see Sect.~\ref{subsect: IS flux calculation}). The central value $\hat{L}$ and its standard deviation $\sigma_{\log L}$ are taken from Ref.~\cite{Weinrich:2020ftb}, and depend on the chosen transport model. In particular, the central values for the halo half-height are $\hat{L} = 4.96^{+2.97}_{-1.77}$~kpc for the BIG model and $\hat{L} = 4.79^{+3.19}_{-1.77}$~kpc for QUAINT. As shown in Refs.~\cite{Donato:2008yx, Genolini:2021doh}, the resulting primary flux is approximately proportional to $L$. Consequently, the uncertainty on $L$ alone induces an asymmetric uncertainty on the predicted primary flux of roughly ${}^{+60\%}_{-35\%}$. This effect dominates over the remaining background errors encoded in the covariance matrix, which are of the order of $20\%$~\cite{Boudaud:2019efq}. Since the ${}^{10}$Be/${}^{9}$Be and ${}^{10}$Be/${}^{7}$Be isotopic ratios act as radioactive clocks, forthcoming high-statistics \textsf{AMS-02} measurements of beryllium isotopes~\cite{berlillium2026} are expected to break current parameter degeneracies and significantly reduce the uncertainties on $L$.

To compute the upper limit on \sigmav{} for a fixed DM mass $m_\chi$, we utilize the profile likelihood ratio (LR) test statistic:
$$ \mathrm{LR}(\sigmav) = -2 \ln \mathcal{L} (L_{\rm min}, \sigmav) + 2 \ln \mathcal{L} ({L}', {\sigmav}') \, , $$
where the first term is evaluated at the halo size $L_{\rm min}$ that maximizes the likelihood for the given \sigmav{}, and the second term is evaluated considering the global best-fit parameters $({L}', {\sigmav}')$. Assuming Wilks' theorem, the LR follows a $\chi^2$ distribution with one degree of freedom. In this way, we determine the $95\%$ CL upper limits by requiring $\mathrm{LR} = 3.84$.

\subsection{Results on \sigmav\ upper limits}
\begin{figure}
    \centering
    \includegraphics[width=1\linewidth]{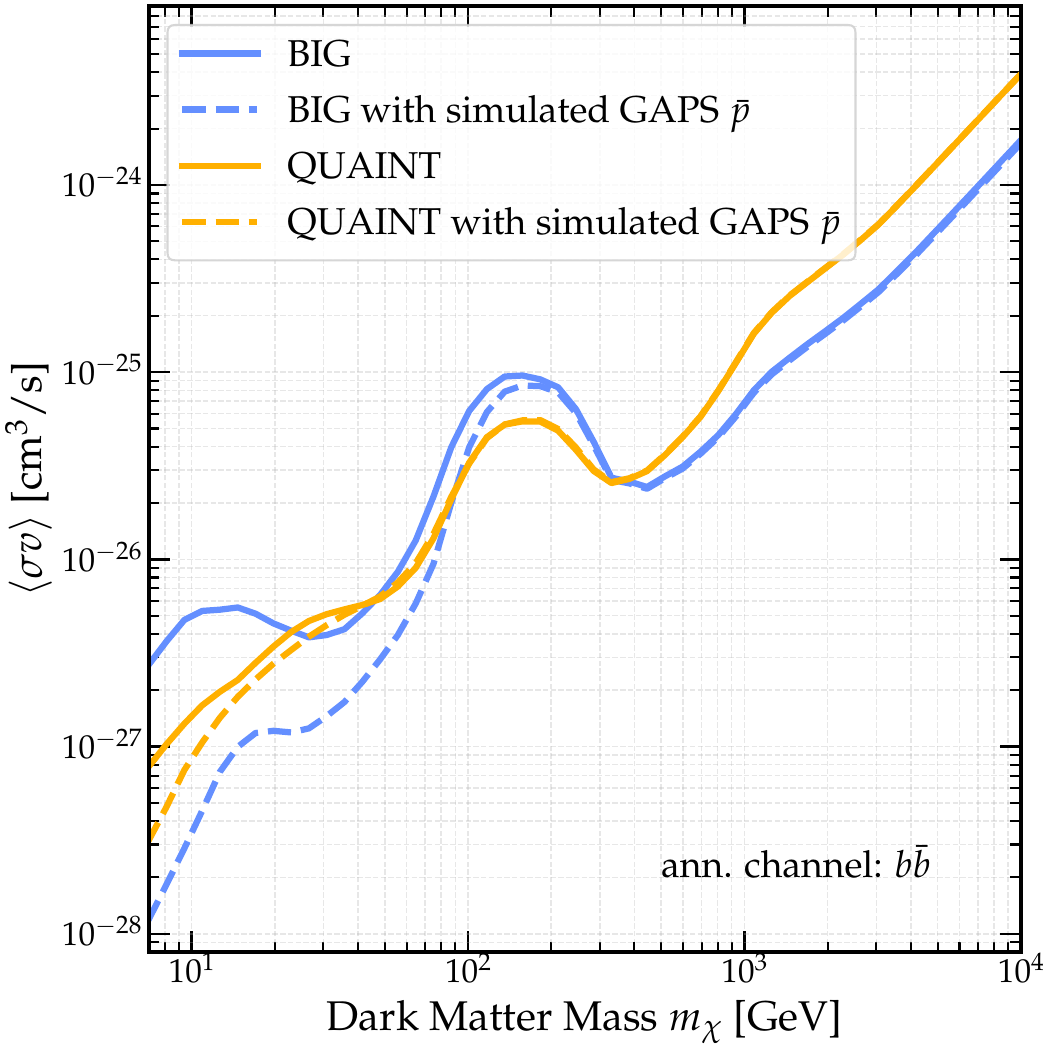}
    \caption{\textbf{$95\%$ CL upper limits on \sigmav{}} as a function of $m_\chi$, assuming the $b\bar{b}$ final state. Solid lines represent the constraints obtained from the analysis of \textsf{AMS-02} \pbar{} data~\cite{AMS:2021nhj} alone. Dashed lines indicate the projected sensitivity gain achieved by including simulated \pbar{} data from \textsf{GAPS}. Results are shown for both the \BIG{} (blue) and \QUAINT{} (yellow) propagation models.}
    \label{fig:sigmav_constraints_BIG_QUAINT_with_GAPS}
\end{figure}

\begin{figure}
    \centering
    \includegraphics[width=1\linewidth]{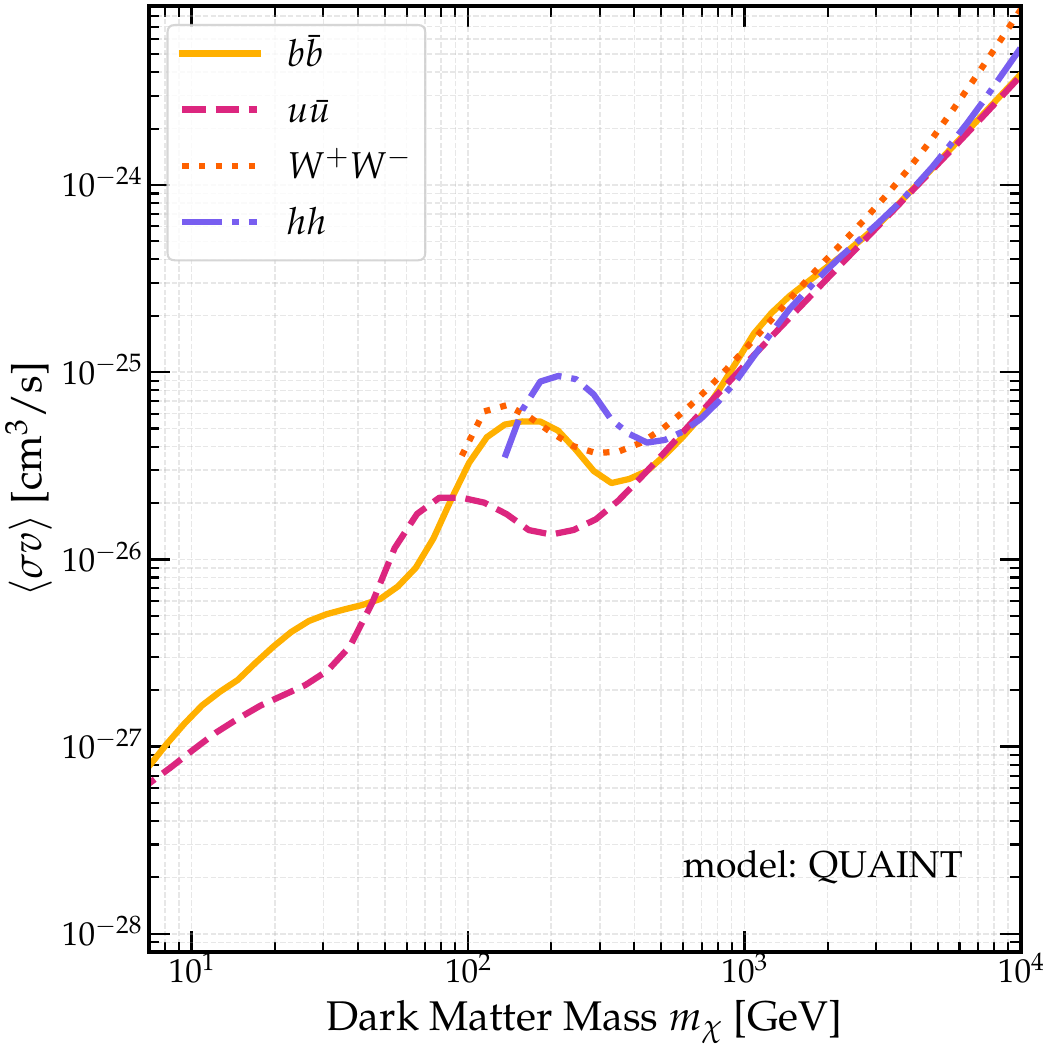}
    \caption{\textbf{$95\%$ CL upper limits on \sigmav{}} as a function of $m_\chi$, obtained with the \QUAINT{} propagation model, assuming the $b\bar{b}$ (solid yellow line), $u\bar{u}$ (dashed magenta), $W^+W^-$ (orange dotted) and $hh$ (purple dot-dashed)  final states.}
    \label{fig:UL_sigmav_dif_channel}
\end{figure}

The resulting constraints are plotted in Fig.~\ref{fig:sigmav_constraints_BIG_QUAINT_with_GAPS}, which displays the upper limits on \sigmav{} as a function of  $m_{\chi}$ ranging from 5\,GeV to 10\,TeV, assuming annihilation into the $b\bar{b}$ channel. The solid lines represent the bounds derived for the \BIG{} and \QUAINT{} propagation models. The dashed lines illustrate the projected impact of incorporating simulated \textsf{GAPS} \pbar{} data, as detailed in Sect.~\ref{subsec:GAPS_pbar}. Additionally, Fig.~\ref{fig:UL_sigmav_dif_channel} presents the corresponding constraints for alternative annihilation channels.

Focusing on the solid lines in Fig.~\ref{fig:sigmav_constraints_BIG_QUAINT_with_GAPS}, the impact of the chosen propagation model on the constraints becomes evident. The \BIG{} model yields less stringent bounds on \sigmav{} for lower DM masses, especially for $m_{\chi} \lesssim 20$\,GeV, whereas the \QUAINT{} model permits a larger DM contribution in the high-mass regime ($m_{\chi} \gtrsim 300$\,GeV).

\begin{figure}[t]
    \centering
    \begin{minipage}{0.48\textwidth}
        \centering
        \includegraphics[width=\textwidth]{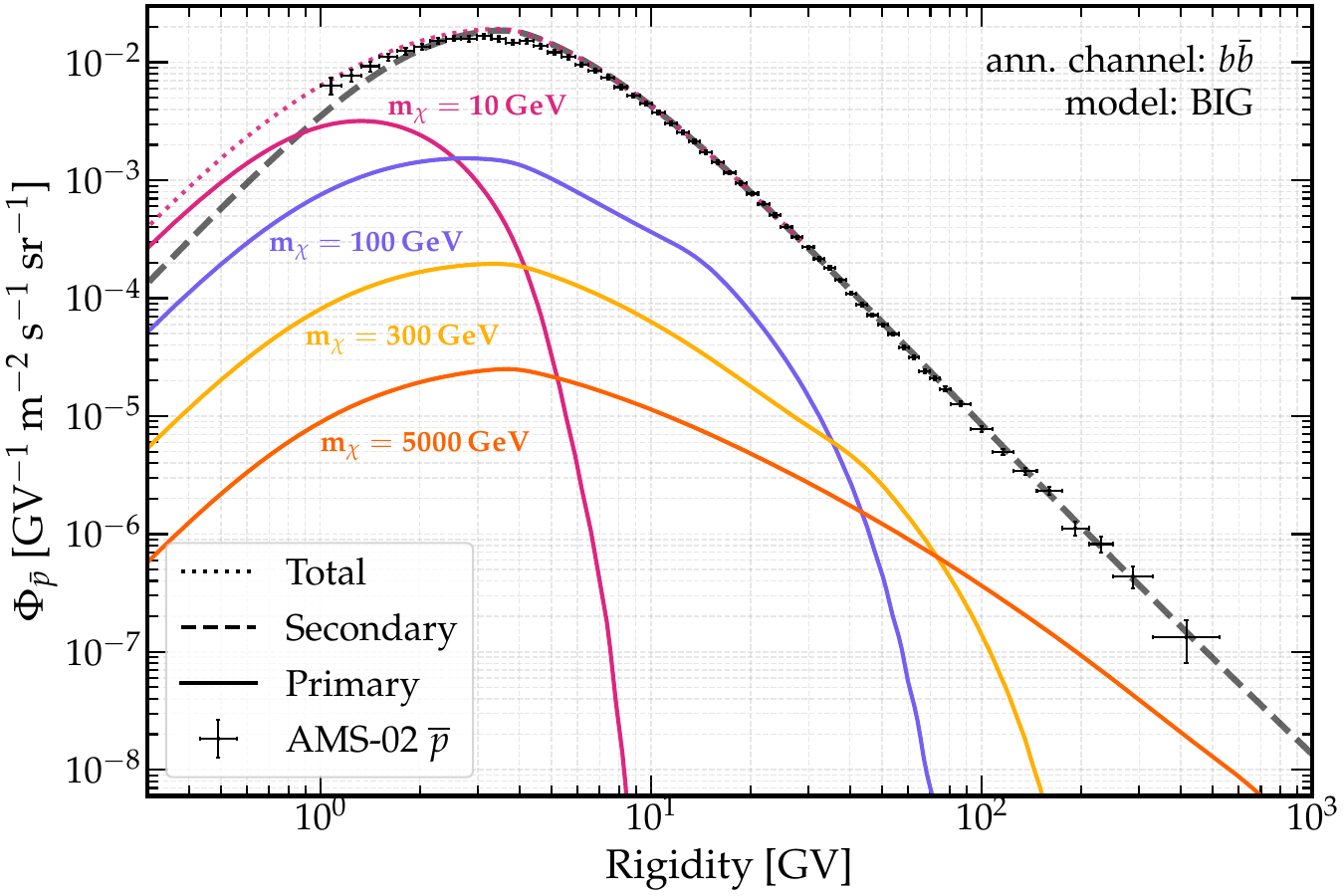}
    \end{minipage}
    \hfill
    \begin{minipage}{0.48\textwidth}
        \centering
        \includegraphics[width=\textwidth]{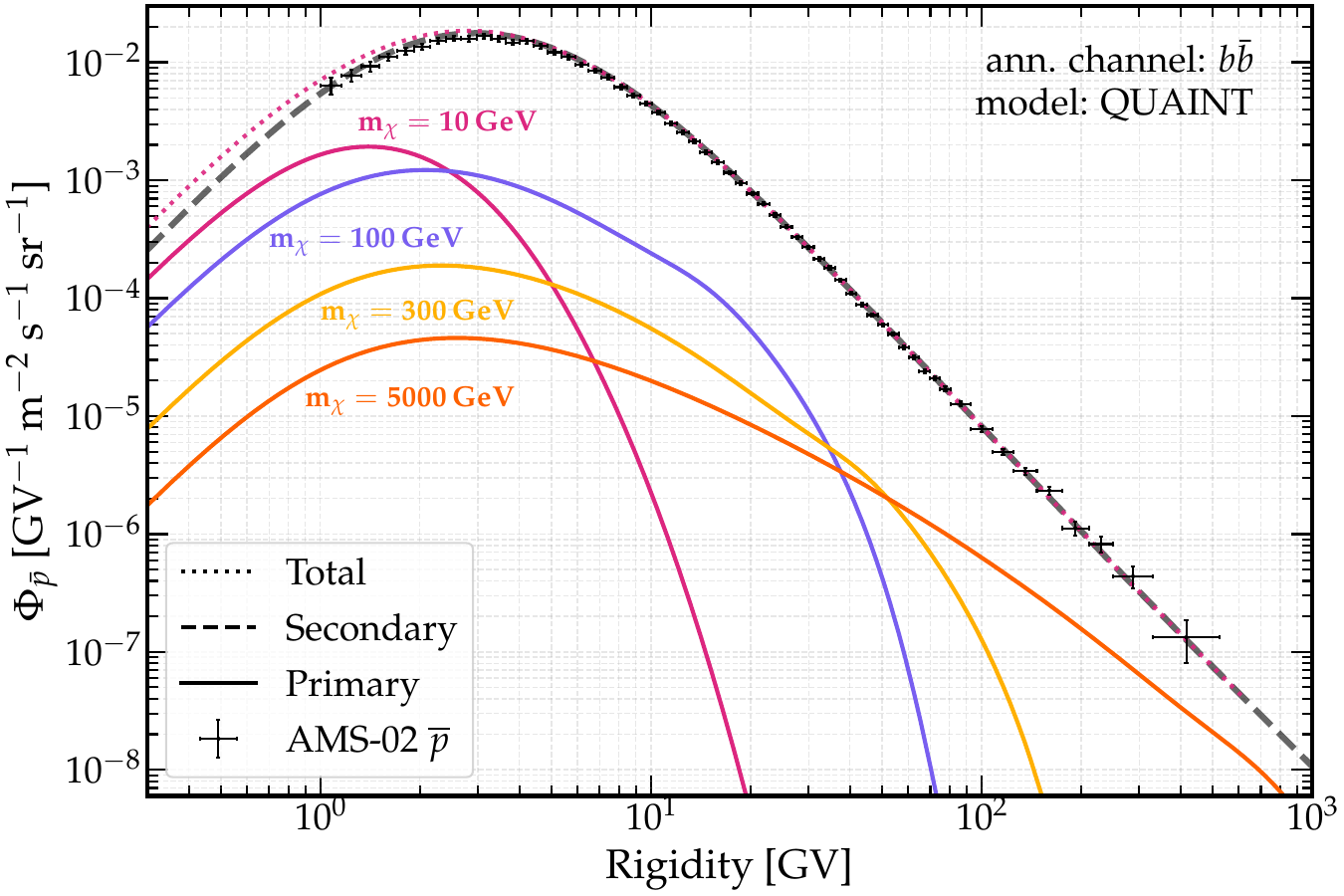}
    \end{minipage}
    \caption{\textbf{TOA antiproton fluxes} evaluated for the \BIG{} (upper panel) and \QUAINT{} (lower panel) propagation models. Theoretical predictions include the secondary flux (gray dashed lines) and the primary contribution from DM annihilation into the $b\bar{b}$ channel (solid lines). The primary fluxes are obtained for the 95\% CL upper limits on \sigmav{}. The DM masses considered are $m_{\chi} = 10$ (magenta), $100$ (purple), $300$ (yellow), and $5000$\,GeV (orange). For the 10\,GeV case, the total flux (primary + secondary) is indicated by the dotted magenta line. \textsf{AMS-02} data~\cite{AMS:2021nhj} are shown for comparison. Solar modulation is treated considering $\phi_{\rm FF} = 0.678$\,GV.}
    \label{fig:UL_pbar_fluxes}
\end{figure}

To understand these spectral features, it is useful to examine the underlying fluxes. Figure~\ref{fig:UL_pbar_fluxes} displays the predicted secondary, primary, and total TOA fluxes of antiprotons for selected DM masses, for both \BIG{} and \QUAINT{} models. For each DM mass, the primary fluxes are computed at the corresponding $95\%$ CL upper limits on \sigmav{} and the associated profile-likelihood value of the diffusion halo half-height $L_{\mathrm{min}}$. Therefore, these curves represent the maximum allowed primary fluxes corresponding to the $95\%$ CL upper limits on \sigmav{}, constrained not to overshoot the \textsf{AMS-02} data. Since the uncertainties on the transport and on the $L$ parameter are already actively profiled out in this limit-setting procedure, overlaying an additional propagation uncertainty band on these specific curves would effectively result in a double-counting of the systematic errors. For an absolute estimation on the propagation uncertainties on primary DM fluxes with sources in the whole halo we refer to Ref. \cite{Genolini:2021doh}.

Examining the physical behavior, the \BIG{} model exhibits a low-rigidity diffusion break around $4$\,GV, which produces a noticeable kink in the flux. This feature is also present in the secondary component, which is steeper at $R\lesssim 4$\,GV than that of the \QUAINT{} model. Consequently, the secondary flux in the \BIG{} model leaves more room for a DM contribution in this rigidity range. This leads to less stringent bounds on \sigmav{} for $m_{\chi} \lesssim 20$\,GeV, as observed in Fig.~\ref{fig:sigmav_constraints_BIG_QUAINT_with_GAPS}. Conversely, the \QUAINT{} model provides tighter constraints  because the corresponding secondary flux follows the \textsf{AMS-02} measurements more closely in the lower rigidity bins. Finally, for masses $m_{\chi} \gtrsim 300$\,GeV, the \QUAINT{} framework becomes less constraining, thus allowing for a slightly larger primary DM contribution. This effect is noticeable for $m_{\chi} = 5000$\,GeV in Fig.~\ref{fig:UL_pbar_fluxes}.

It is important to emphasize that the secondary fluxes are derived prior to the present analysis: they are \textit{predicted} and intentionally \textit{not fitted} to the antiproton data. As discussed in Sect.~\ref{subsect: IS flux calculation}, neither model is statistically strongly preferred by current nuclei or antiproton data. Therefore, we utilize both to bracket the systematic uncertainty associated with the transport setup.

\subsection{Discussion on spectral features effects on \sigmav{} upper limits}
\label{subsec:sigmav_UL_visual_expl}
\begin{figure*}
    \centering
    \includegraphics[width=0.75\linewidth]{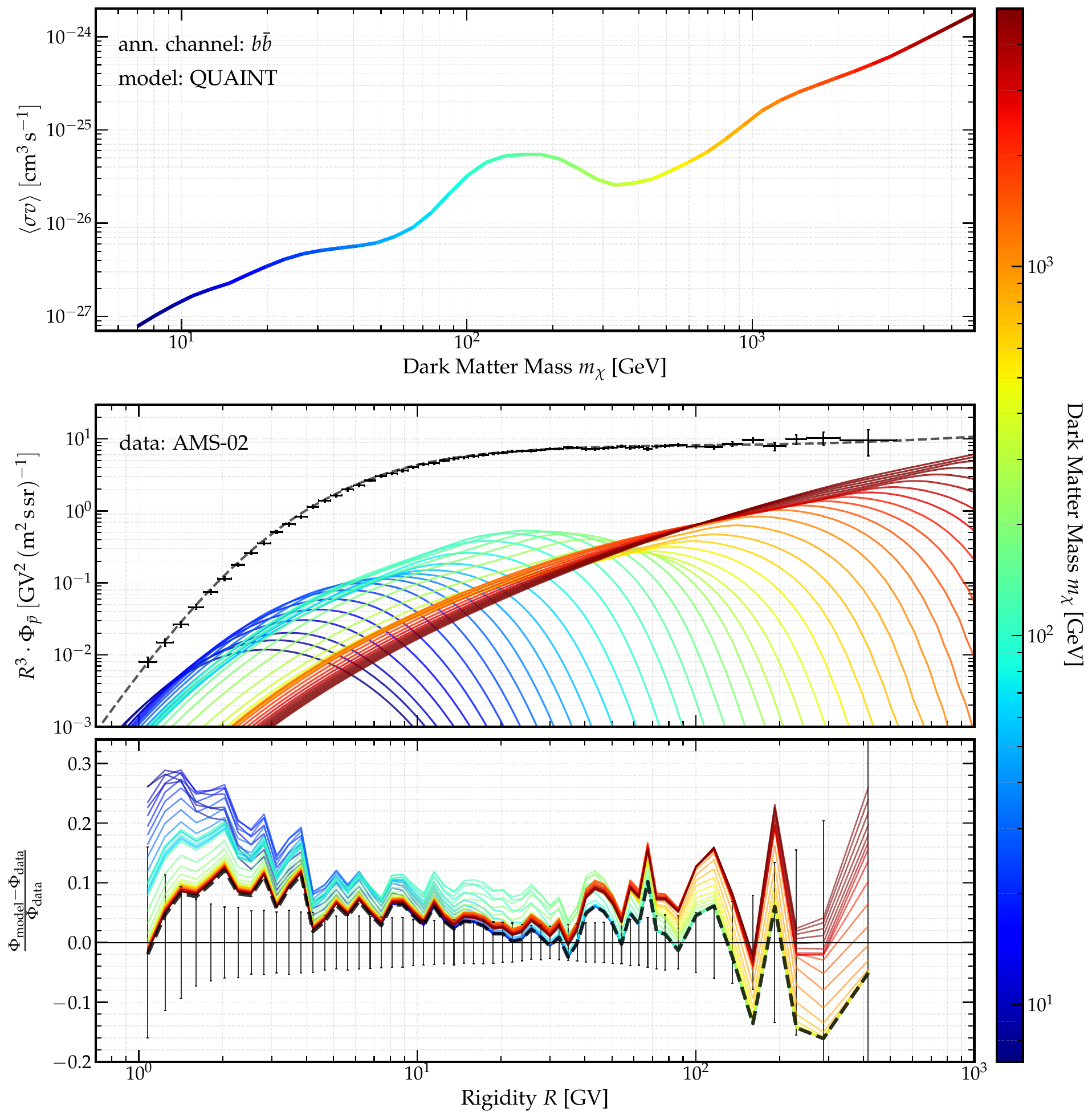}
    \caption{
            \textbf{Top panel}: upper limits on \sigmav{} obtained in our antiproton analysis for the \QUAINT{} transport model, as a function of the DM mass $m_{\chi}$.
            \textbf{Middle panel}: primary antiproton fluxes for various DM masses ranging from 7 to 5000\,GeV (see the color bar on the right for the mass-color correspondence). These fluxes have been computed assuming the upper limit value of \sigmav{} corresponding to each mass, as shown in the top panel. The fluxes are rescaled by a factor of $R^3$. The gray dashed line represents the secondary background reference flux (predicted, not fitted to the data). Black markers indicate the \textsf{AMS-02} data \cite{AMS:2021nhj}.
            \textbf{Bottom panel}: residuals from the comparison between the total fluxes (secondary + primary at the upper limit level) and the \textsf{AMS-02} data. The black dashed line represents the residuals of the secondary reference flux alone. These residuals have been directly calculated without including the effect of the data covariance matrix and, for visual clarity, without showing the uncertainties on the secondary fluxes.
            }
    \label{fig: QUAINT constraints visual breakdown}
\end{figure*}

To discuss more in detail the spectral features effects on the determination of the \sigmav{} constraints, Fig.~\ref{fig: QUAINT constraints visual breakdown} shows a visual breakdown of how the exclusion limits are formed. Although the bottom panel simplifies the analysis by neglecting the full covariance matrix of the data, the comparison of these three panels effectively highlights the correlation between the spectral features of the primary and secondary \pbar{} fluxes and the resulting shape of the constraints. In particular, it helps clarifying the origin of the weakening of the limits at approximately $80 \text{--} 250$\,GeV in $m_{\chi}$, a feature also present in other works where correlations in \textsf{AMS-02} data were not considered, such as Ref.~\cite{Balan:2023lwg}.
The first two panels visually show the correspondence between the shape of the constraint curve (top panel) and the position of the peaks of the primary fluxes rescaled by $R^3$ (middle panel), computed using the upper limit values of \sigmav{}\ and the associated value of $L_{\rm min}$.
From the plot of the residuals (third panel), it is possible to visually infer the origin of the spectral shape of the constraints, which results from the interplay between the position of the primary flux peaks and the shape of the predicted secondary background. Specifically, the weakening of the constraints at $m_{\chi} \approx 80 \text{--} 250$\,GeV in the top panel corresponds to a region in the third panel at $R \approx 10 \text{--} 40$\,GV, aligning more closely  (within the $1\sigma$ experimental uncertainty) to the \textsf{AMS-02} mean value, thus leaving room for a larger DM contribution in the upper limit analysis.

In conclusion, although the exact correspondence between residuals and limits is inevitably smoothed by data correlations and secondary uncertainties, this visualization successfully captures the most significant qualitative effects, providing an effective insight into the physical origin of the constraints' shape.

\subsection{Result with simulated \pbar{} GAPS data}
\label{subsec:GAPS_pbar}

Having established the constraints from current \textsf{AMS-02} data, we explore the discovery reach of the upcoming \textsf{GAPS} experiment. \textsf{GAPS} specifically targets the sub-GV antiproton window, where the secondary background is significantly suppressed, thereby maximizing the sensitivity to light DM signals. To quantitatively gauge the impact of these future measurements on the \sigmav{} constraints, we construct an illustrative scenario projecting the instrument's sensitivity onto the \BIG{} and \QUAINT{} background models, explained as follows. 

We extract the projected energy positions, flux values, and relative errors of simulated \textsf{GAPS} data from Fig.~9 in Ref.~\cite{GAPS:2022ncd}. Following the Asimov dataset approach~\cite{Cowan_2011}, we generate a single representative mock dataset for each transport setup by forcing the \textsf{GAPS} data points to exactly match the median expected values of our secondary backgrounds. We treat the projected \textsf{GAPS} relative uncertainties as uncorrelated statistical errors. By neglecting correlated systematic uncertainties, this procedure yields an optimistic projection of the exclusion limits. Furthermore, because of the ambiguity regarding the exact solar modulation conditions during the upcoming \textsf{GAPS} flights, we adopt the \textsf{AMS-02} modulation levels as a standardized baseline. 

The results of this exercise are shown as dashed lines in Fig.~\ref{fig:sigmav_constraints_BIG_QUAINT_with_GAPS}. For the \BIG{} model, the addition of these simulated low-energy data can significantly improve the constraints on \sigmav{} for low DM masses, even more than one order of magnitude. This is due to the steeper secondary flux shape in the sub-GV region, which results in stronger constraints. At the same time, the improvement in the DM bound by using the extrapolated background can also be rephrased by saying that for DM masses $\lesssim50$~GeV, \textsf{GAPS} potentially has an excellent chance of detecting a DM signal in the antiproton channel. Conversely, for the \QUAINT{} model, the effect is milder since the predicted background flux is higher.

\begin{figure}
    \centering
    \includegraphics[width=1\linewidth]{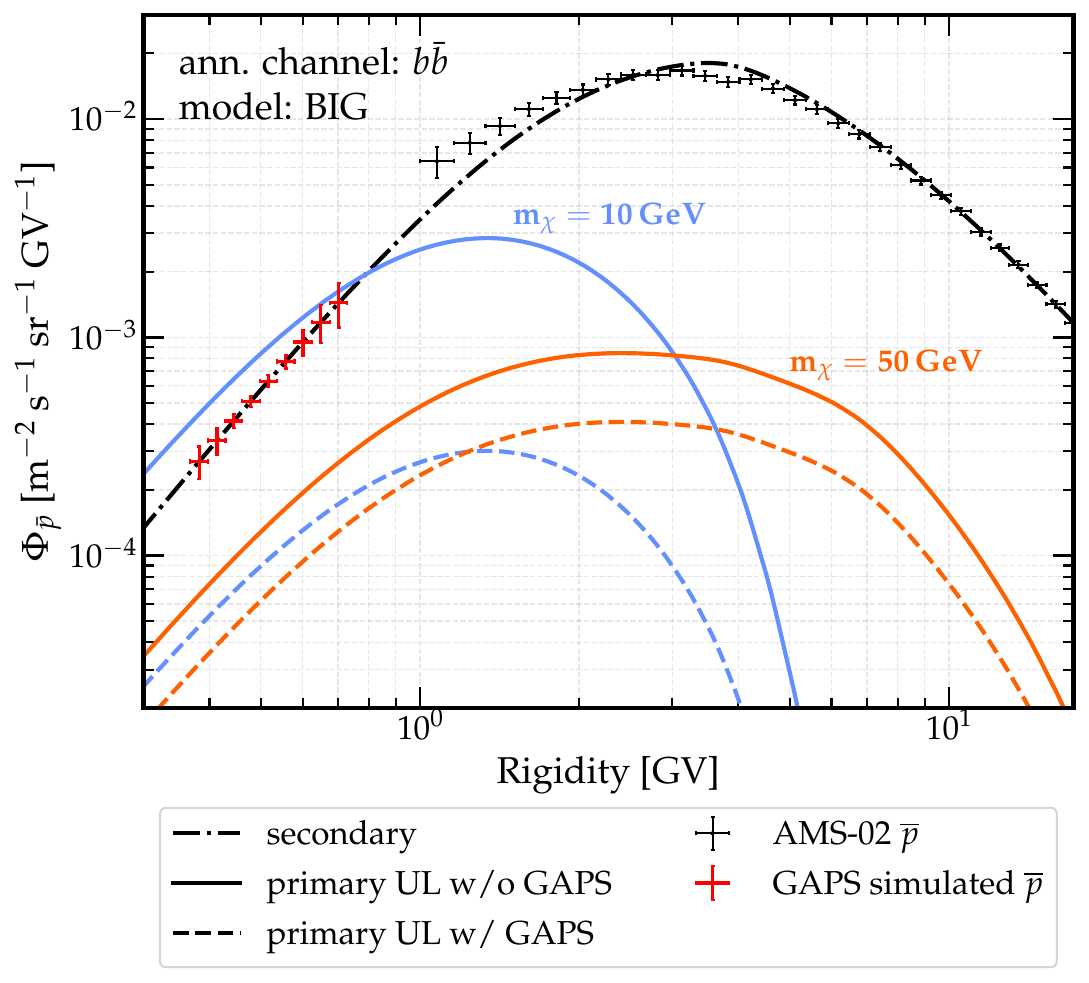}
    \caption{\textbf{Effect of mock \textsf{GAPS} data on the maximum allowed primary TOA antiproton fluxes}, evaluated for the \BIG{} propagation model. Solid lines represent the maximum primary fluxes derived considering
     \textsf{AMS-02} data alone, whereas dashed lines illustrate the constraints obtained considering the mock \textsf{GAPS} dataset in the \sigmav{} bound computation. 
     The plot displays the mock \textsf{GAPS} data (red crosses) superimposed on the predicted secondary background (dash-dotted black line), alongside the \textsf{AMS-02} data~\cite{AMS:2021nhj} (black crosses). The primary fluxes are computed for two representative masses of 10 and 50\,GeV annihilating into $b\bar{b}$, and at the 95\% CL upper limits on \sigmav{} as in  Fig.~\ref{fig:sigmav_constraints_BIG_QUAINT_with_GAPS}. }
    \label{fig:simulated_GAPS_pbar}
\end{figure}

To visualize this impact, Fig.~\ref{fig:simulated_GAPS_pbar} displays the mock \textsf{GAPS} data projected onto the \BIG{} secondary background, alongside the 2011--2018 \textsf{AMS-02} measurements~\cite{AMS:2021nhj}. The plot clearly illustrates how the addition of sub-GV data restricts the maximum allowed primary DM fluxes. This constraining power is particularly pronounced for lighter DM candidates, such as the $m_{\chi} = 10$\,GeV case, because their annihilation spectra peak closer to the energy window of \textsf{GAPS}.

\section{Results on antideuteron fluxes}
\label{sec:results_antid}
Having established robust upper limits on \sigmav{} from the antiproton channel, we now turn to antideuterons. We aim to determine whether a primary \dbar{} signal could still fall within the reach of current or near-future detectors without violating the stringent \pbar{} bounds derived above. 

\begin{figure}[t]
    \centering
    \begin{minipage}{0.48\textwidth}
        \centering
        \includegraphics[width=\textwidth]{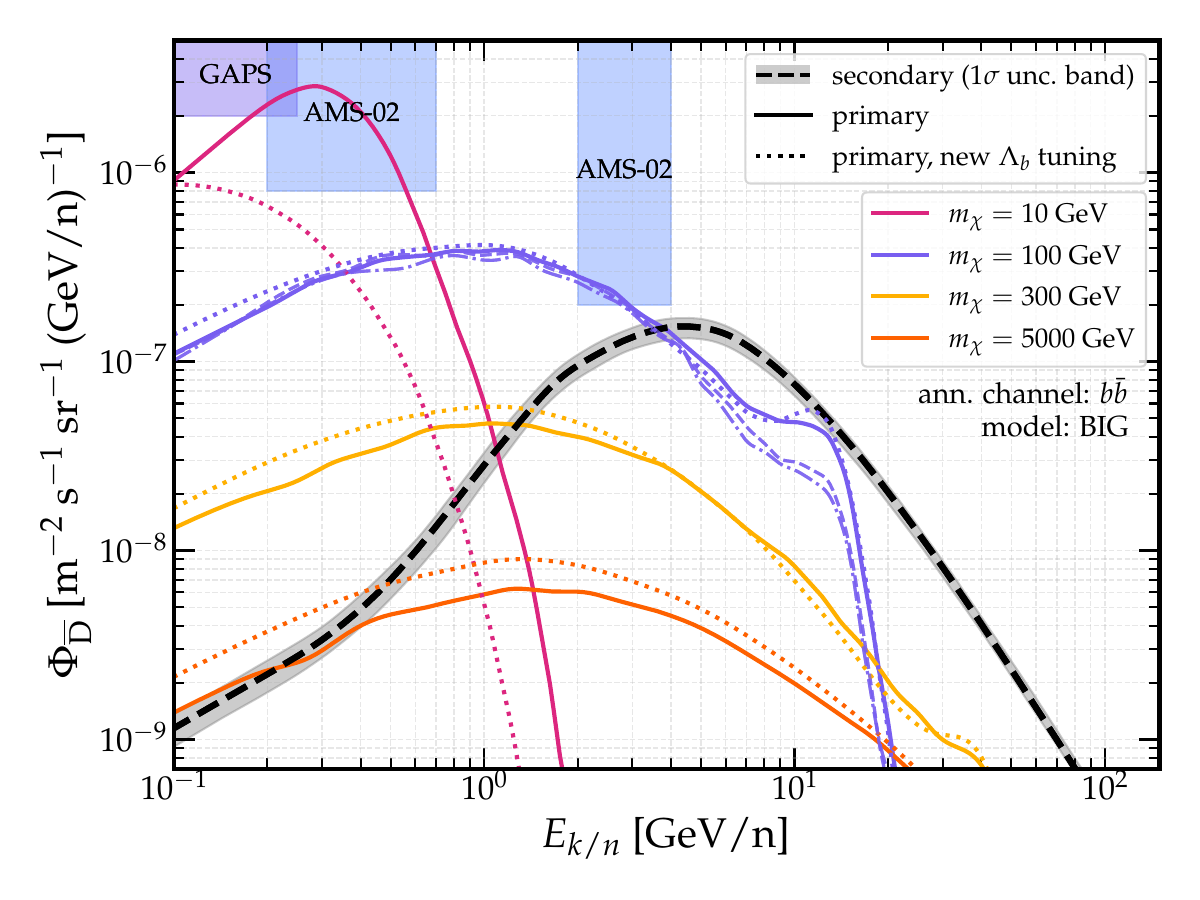}
    \end{minipage}
    \hfill
    \begin{minipage}{0.48\textwidth}
        \centering
        \includegraphics[width=\textwidth]{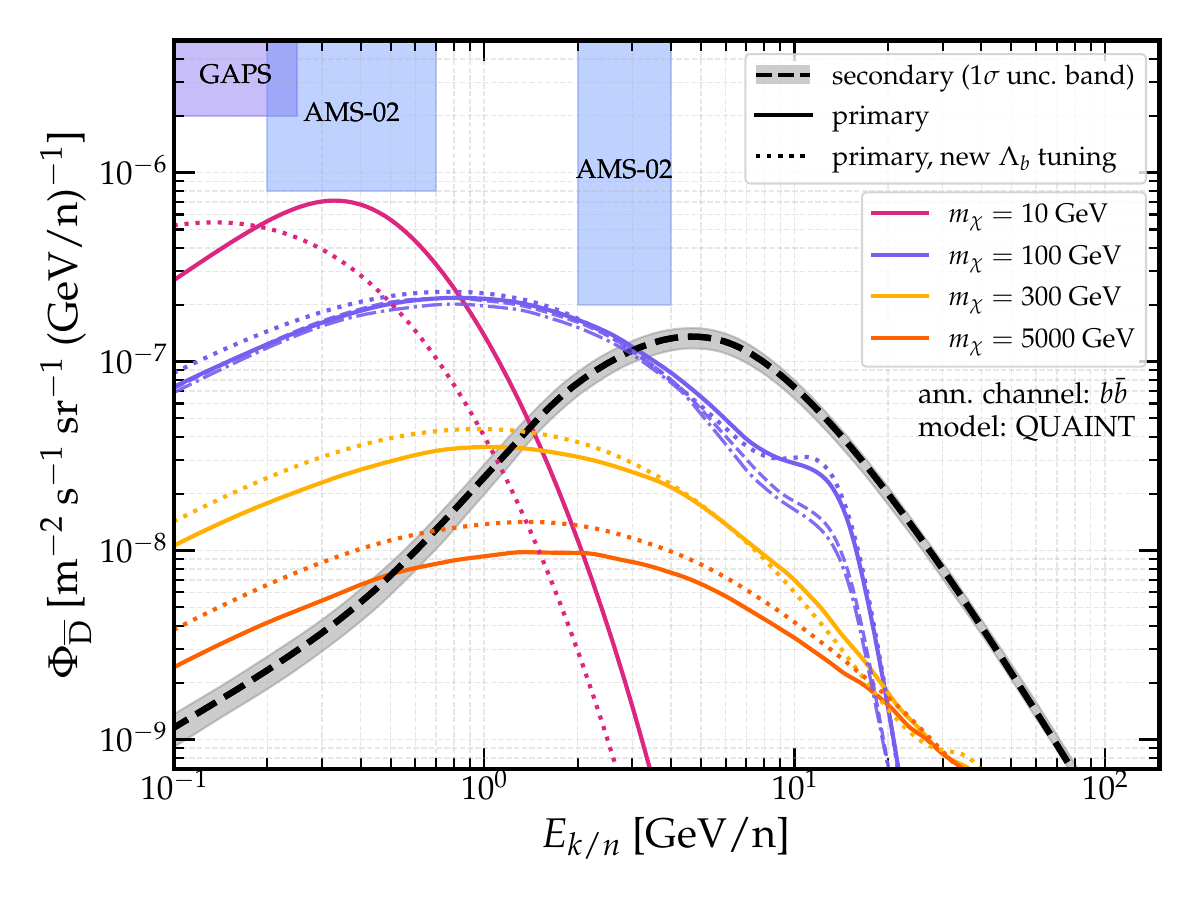}
    \end{minipage}
    \caption{\textbf{TOA antideuteron fluxes.} The plot displays the expected secondary background flux (black dashed line) alongside primary fluxes from DM annihilation, considering the ${b\bar{b}}$ annihilation channel. The primary components are shown for both the standard \texttt{CosmiXs} spectra (solid lines) and the updated $\Lambda_b$ tuning (dotted lines). Unless otherwise specified, all primary fluxes are computed using the Argonne coalescence model. For the $m_\chi = 100$\,GeV candidate, the predictions from the Gaussian (dashed purple line) and $\Delta p + \Delta r$ (dash-dotted purple line) approaches are also shown for comparison. All fluxes are evaluated at a solar modulation potential $\phi_{\rm FF}=750$\,MV. The primary signals are computed using the 95\% CL upper limits on \sigmav{} derived from our \pbar{} analysis. The gray band indicates the theoretical uncertainty on the secondary flux. The upper and lower panels present the results for the \BIG{} and \QUAINT{} propagation models, respectively.}
    \label{fig:UL_dbar_fluxes}
\end{figure}

\begin{figure}
    \centering
    \includegraphics[width=1\linewidth]{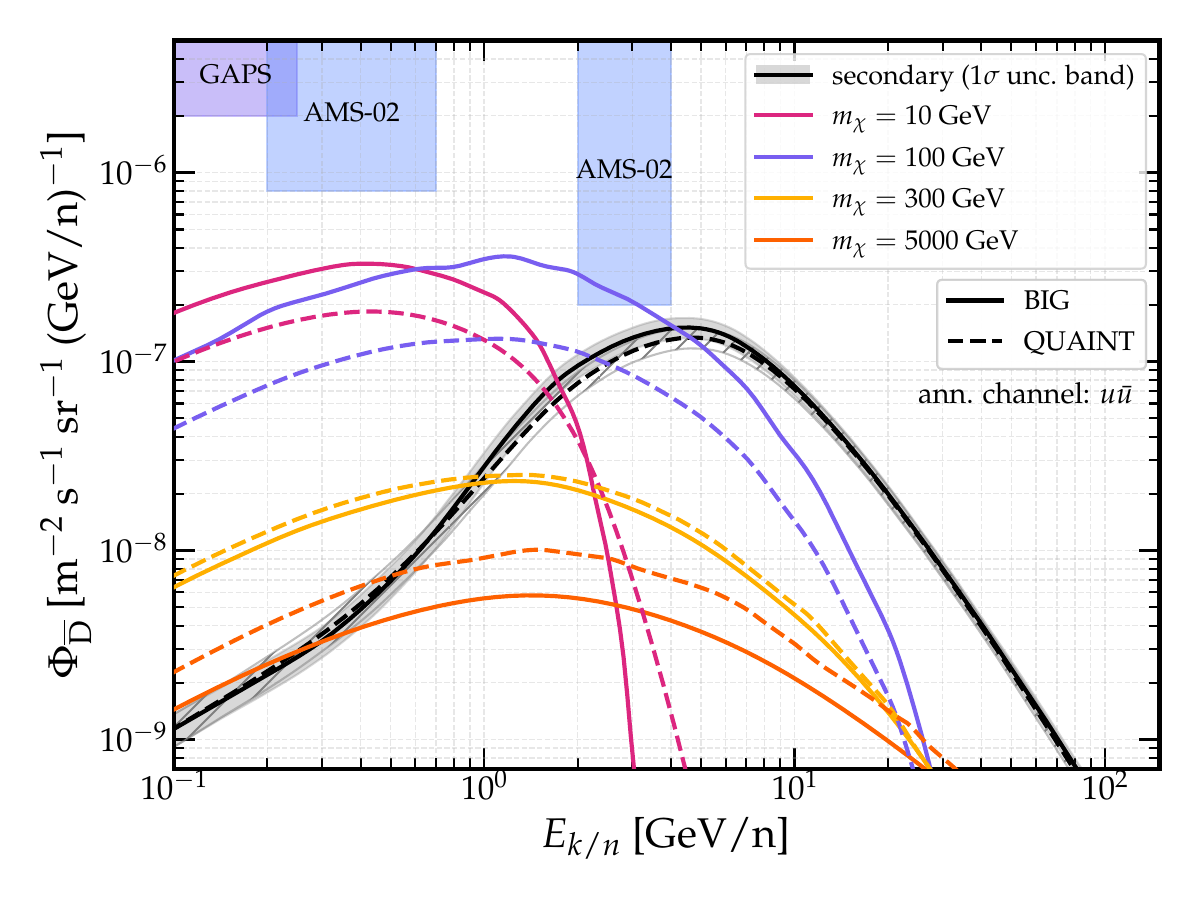}
    \caption{\textbf{TOA antideuteron fluxes.} The plot displays the expected secondary background fluxes (black lines) alongside primary fluxes from DM annihilation, considering the ${u\bar{u}}$ annihilation channel. The fluxes are shown for both the \BIG{} (solid lines) and the \QUAINT{} (dotted lines) propagation models. All fluxes are evaluated at a solar modulation potential $\phi_{\rm FF}=750$\,MV. The primary signals are computed using the 95\% CL upper limits on \sigmav{} derived from our \pbar{} analysis and employing the the standard \texttt{CosmiXs} spectra. The $1\sigma$ theoretical uncertainties on the secondary fluxes are represented by the gray bands: solid-filled for the \BIG{} model and hatched for the \QUAINT{} model.}
    \label{fig:u_ubar}
\end{figure}

To address this, we compute the maximum allowed primary \dbar{} fluxes. For each DM mass $m_{\chi}$, we adopt the $95\%$ CL upper limit on \sigmav{} alongside the corresponding profile-likelihood value of the halo half-height $L_{\mathrm{min}}$. Given the uncertainty regarding the exact solar activity during the projected \textsf{GAPS} flights, we adopt a reference Force-Field modulation potential of $\phi_{\rm FF} = 750$\,MV for the following evaluations. Figure~\ref{fig:UL_dbar_fluxes} (\ref{fig:u_ubar}) illustrates these upper-bound primary antideuteron fluxes for selected DM masses for the ${b\bar{b}}$ (${u\bar{u}}$) annihilation channel, displaying the maximum expected signal against the secondary background. The plots also include the projected experimental sensitivities: the expected reach of three balloon flights for \textsf{GAPS}~\cite{Aramaki_2016}, and the projected 2030 flux sensitivity for \textsf{AMS-02}, which accounts for its planned acceptance upgrade~\cite{2008ICRC....4..765C,Oliva:JENAA:2024}.

Focusing first on the secondary background, the flux peaks at approximately $4$\,GeV/$n$, approaching the sensitivities of \textsf{AMS-02} and \textsf{GAPS} for both propagation models. The plots display the theoretical uncertainty band, whose computation is detailed in Sect.~\ref{subsect: IS flux calculation}. To bracket the systematic uncertainty arising from the coalescence process, we evaluated the envelope of the predictions obtained from different coalescence models (namely, the $\Delta p + \Delta r$, Gaussian-Wigner, and Argonne-Wigner models), adopting the Gaussian model as the central baseline.\\ 

Considering Fig.~\ref{fig:UL_dbar_fluxes}, for both propagation models the maximum allowed primary signal exhibits a strong mass dependence: the lower the DM mass, the larger the maximum predicted excess over the secondary background at sub-GeV energies. For instance, in the case of an $10$\,GeV DM candidate, the primary flux can exceed the background by up to three orders of magnitude. This clearly reaffirms the role of low-energy antideuterons as an almost background-free channel for indirect DM detection~\cite{Donato:1999gy}.

Comparing these maximum allowed primary fluxes with experimental sensitivities reveals a significant dependence on the underlying propagation framework. Under the \BIG{} model, which yields less stringent \sigmav{} constraints for DM masses $m_\chi \lesssim 20$\,GeV, the primary spectra can reach the projected \textsf{GAPS} sensitivity for very light DM candidates. Conversely, the \QUAINT{} model yields tighter low-mass constraints, suppressing the maximum allowed fluxes below the \textsf{GAPS} detection threshold across the entire probed mass range.

The plots show the impact of the updated $\Lambda_b$ tuning, discussed in Sect.~\ref{sec:prim_production}, on the primary antideuteron fluxes. This modification primarily affects the spectra generated by light DM candidates, whereas the impact on heavier masses is substantially milder. This effect has important implications for indirect detection prospects, as the updated tuning suppresses the expected signal at masses $\lesssim 10$\,GeV, rendering it inaccessible to both \textsf{GAPS} and \textsf{AMS-02}.

For the $m_\chi = 100$\,GeV candidate, the plots also illustrate the effect of the different coalescence approaches (i.e., the Gaussian and $\Delta p + \Delta r$ models, in addition to the default Argonne baseline). As evident from the figures and further detailed in Ref.~\cite{DiMauro:2024kml}, the systematic spread introduced by the choice of coalescence model is no longer a dominant source of theoretical uncertainty.
\\

In Fig.~\ref{fig:u_ubar} we show the productions for the \dbar{} generated from the $u\bar{u}$ channel. The main differences from the $b\bar{b}$ channel are the lower predicted fluxes for the lightest DM masses, which consequently fall below the projected \textsf{GAPS} sensitivity. This is due to the absence of \dbar{} contribution from $\Lambda_b$ decays for the $u\bar{u}$ channel compared to the $b\bar{b}$ case, as also noticeable in the corresponding \dbar{} production spectra in Ref.~\cite{DiMauro:2024kml}.\\

\begin{figure}[t]
    \centering
    \begin{minipage}{0.48\textwidth}
        \centering
        \includegraphics[width=\textwidth]{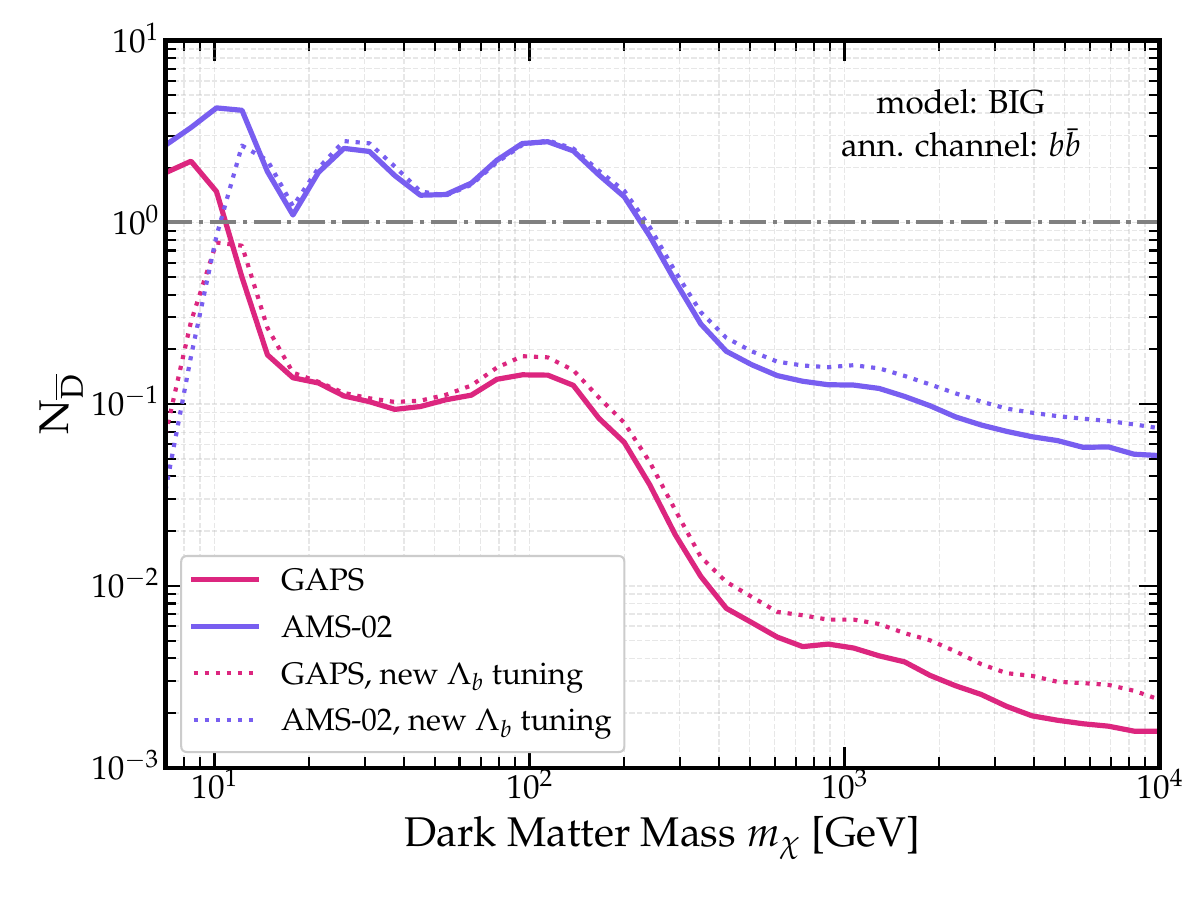}
    \end{minipage}
    \hfill
    \begin{minipage}{0.48\textwidth}
        \centering
        \includegraphics[width=\textwidth]{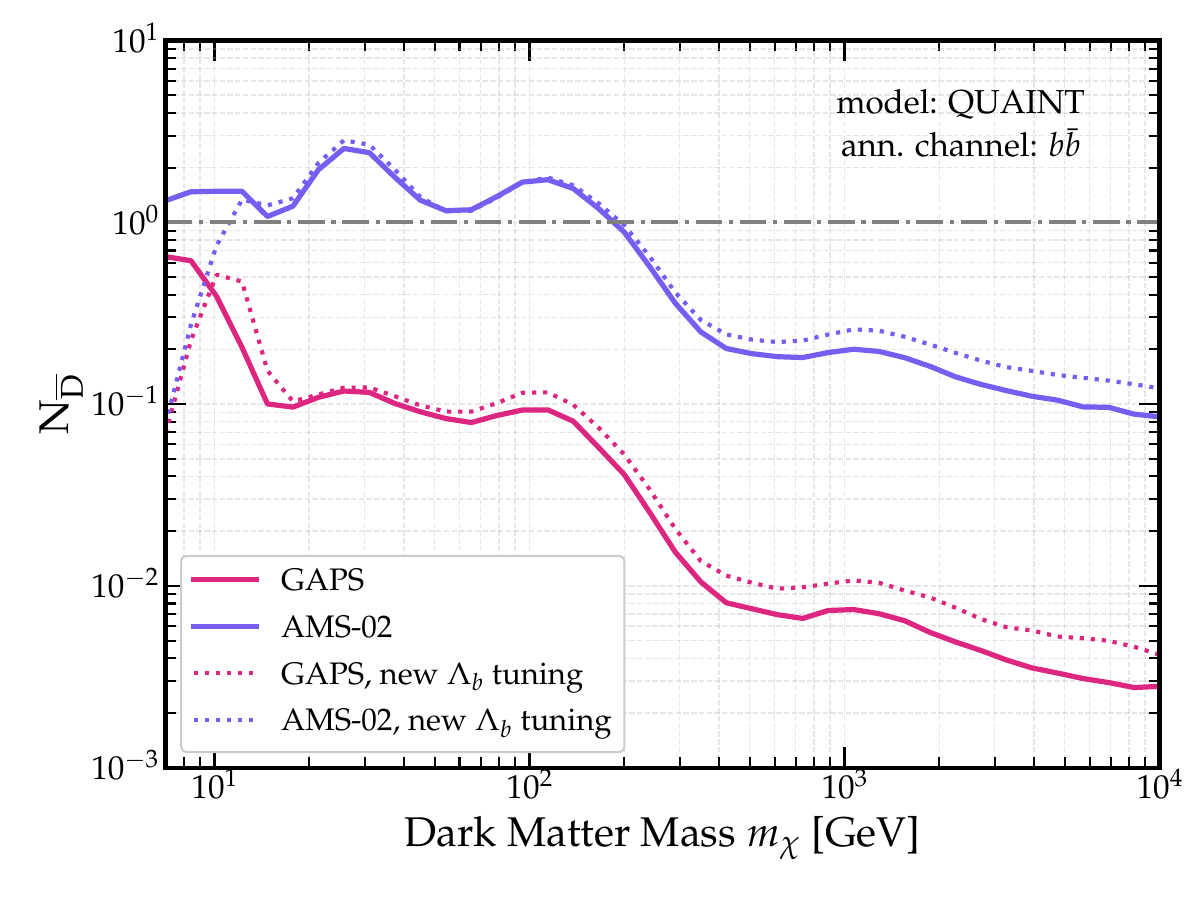}
    \end{minipage}
   \caption{\textbf{Maximum expected number of detectable primary antideuterons ($N_{\overline{\mathrm{D}}}$)} for the \textsf{GAPS} (magenta lines) and \textsf{AMS-02} (purple lines) experiments as a function of the DM mass $m_{\chi}$, computed considering $\phi_{\rm FF}=750$\,MV. The solid lines indicate the yields computed using the standard \texttt{CosmiXs} $\Lambda_b$ tuning, while the dotted lines represent the results obtained with the updated $\Lambda_b$ tuning. The upper and lower panels display the projections for the \BIG{} and \QUAINT{} propagation models, respectively.}
    \label{fig:number_dbar_expected}
\end{figure}

The \dbar{} detection prospects are explicitly quantified in Fig.~\ref{fig:number_dbar_expected}, which displays the maximum number of primary antideuterons that \textsf{GAPS} and \textsf{AMS-02} are predicted to detect, assuming the current \pbar{}-derived constraints and considering the $b\bar{b}$ annihilation channel. The calculation of these expected event yields follows the methodology outlined in Ref.~\cite{dimauro2026enhancedcosmicrayantinucleifluxes}. 
Specifically, the expected number of events is derived by integrating the theoretical primary flux over the kinetic energy window where the experiment is sensitive. This integrated flux is then divided by the energy bin width and normalized by the corresponding experimental flux sensitivity. Finally, the result is multiplied by the baseline number of signal events defining that specific sensitivity threshold. As detailed therein, the flux sensitivities for antideuterons for both \textsf{AMS-02} and \textsf{GAPS} are calibrated to correspond to 2 expected detection events. 

Analyzing the projected yields, under the \BIG{} propagation framework, \textsf{GAPS} is projected to detect at least one primary antideuteron only for light DM candidates with $m_{\chi} \lesssim 10$\,GeV. In contrast, the \QUAINT{} model yields a maximum signal that remains completely inaccessible to \textsf{GAPS} across the entire mass range. For \textsf{AMS-02}, however, both propagation models allow for the potential detection of a primary antideuteron signal for DM masses up to $m_{\chi} \approx 200$\,GeV. 

Incorporating the updated $\Lambda_b$ tuning reveals a strong suppression of the detection yields for DM masses $m_\chi \lesssim 15$\,GeV. Specifically, for masses $m_\chi \lesssim 10$\,GeV, this attenuation pushes the primary signal entirely below the 1-particle detection thresholds of both \textsf{AMS-02} and \textsf{GAPS}. Consequently, since \textsf{GAPS} was exclusively sensitive to this specific low-mass window, this tuning renders the primary signal completely inaccessible to the \textsf{GAPS} experiment across the entire probed mass range. 

Applying the same event yield methodology to the secondary astrophysical background, the expected number of events at \textsf{AMS-02} is evaluated to be $1.12 \pm 0.15$ events under the \QUAINT{} model and $1.31 \pm 0.18$ events under the \BIG{} framework. In contrast, the expected background at \textsf{GAPS} is suppressed by nearly three orders of magnitude relative to the detection threshold, firmly placing the experiment in a virtually background-free regime. \\

\begin{figure}[t]
    \centering
    \begin{minipage}{0.48\textwidth}
        \centering
        \includegraphics[width=\textwidth]{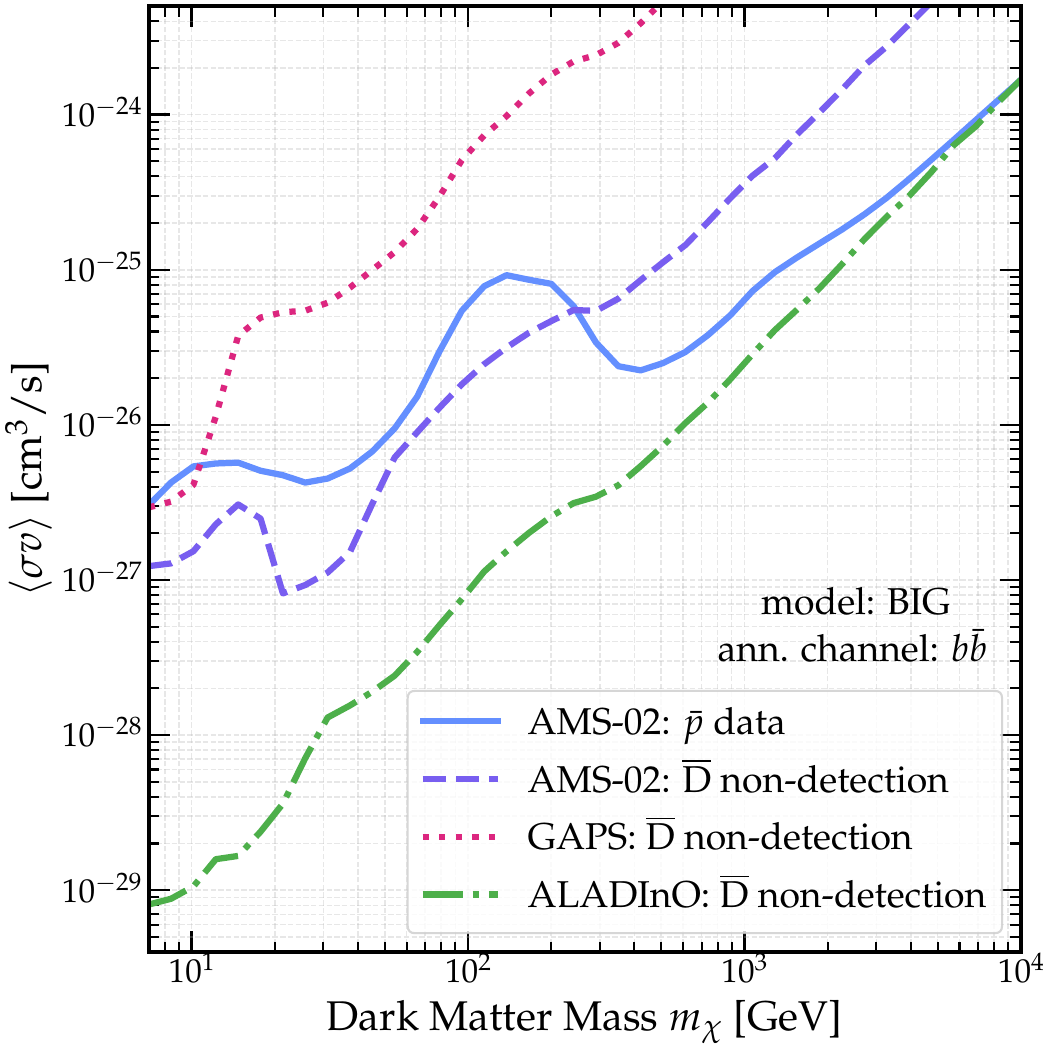}
    \end{minipage}
    \hfill
    \vspace{5pt}
    \begin{minipage}{0.48\textwidth}
        \centering
        \includegraphics[width=\textwidth]{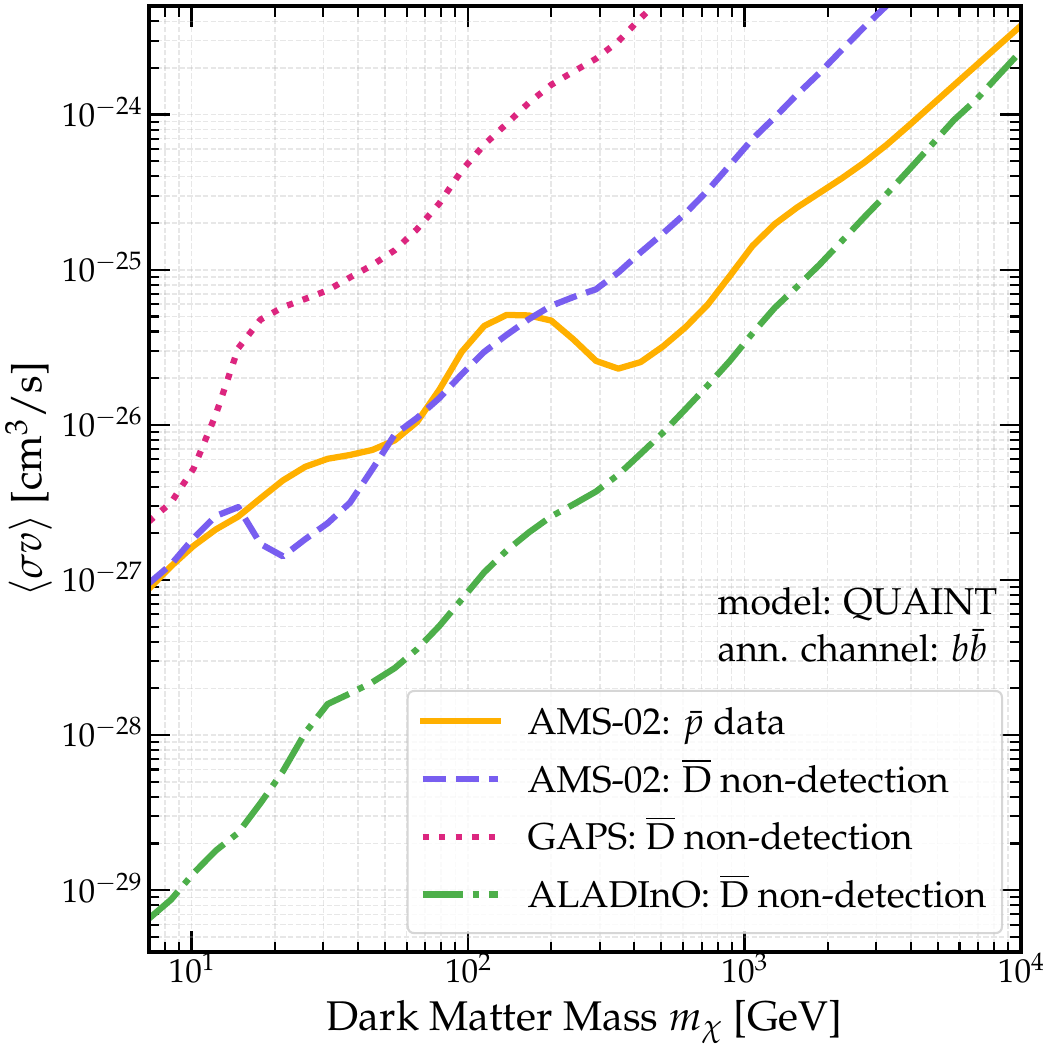}
    \end{minipage}
    \caption{\textbf{95\% CL upper limits on \sigmav{} as a function of $m_{\chi}$}, assuming the $b\bar{b}$ final state. The solid lines represent the current constraints derived from the \textsf{AMS-02} antiproton data. The other curves illustrate the projected upper limits assuming a null detection of primary antideuterons by current and future experiments: \textsf{AMS-02} (dashed purple), \textsf{GAPS} (dotted magenta), and \textsf{ALADInO} (dash-dotted green). Results are displayed for the \textsf{BIG} (upper panel) and \QUAINT{} (lower panel) propagation models.}
    \label{fig:UL_sigmav_dbar_non_detection}
\end{figure}

Since \textsf{AMS-02} and \textsf{GAPS} are projected to be sensitive to primary antideuterons, the absence of a detection can be exploited inversely to constrain the DM parameter space, yielding prospective upper limits on \sigmav{}. To address this, Figure~\ref{fig:UL_sigmav_dbar_non_detection} presents the upper limits derived under the assumption of a null detection of antideuterons by each of these instruments independently. To compute these projected bounds, we adopted a single-bin approach: for each mass, the annihilation cross-section is rescaled such that the predicted total antideuteron flux reaches, but does not exceed, the sensitivity threshold of the respective experiment in its most constraining energy bin.

The results indicate that a confirmed non-detection by \textsf{AMS-02} would improve the current \pbar{}-derived constraints for $m_{\chi} \lesssim 200$\,GeV by up to a factor of $\approx 5$ under the \BIG{} model and $\approx 3$ under \QUAINT{}, exhibiting a non-monotonic mass dependence. Conversely, a null result from \textsf{GAPS} would only tighten the existing limits in the very low mass regime, and strictly under the assumptions of the \textsf{BIG} model. Incorporating the projected capabilities of the proposed \textsf{ALADInO} experiment~\cite{instruments6020019}, a null detection under this simplified framework would drastically reduce the available DM parameter space, effectively constraining the thermal relic cross-section up to $m_{\chi} \approx 1000$\,GeV.

\section{Results on Antihelium  fluxes}
\label{sec:results_antiHe}

As a final step, we assess the detectability of a potential $^3\overline{\mathrm{He}}$ signal by computing the corresponding fluxes under the upper limits on $\langle\sigma v\rangle$ derived from $\overline{\mathrm{p}}$ data, and comparing them with the projected sensitivities of \textsf{AMS-02}~\cite{dimauro2026enhancedcosmicrayantinucleifluxes}, \textsf{GAPS}~\cite{SAFFOLD2021102580}, and \textsf{ALADInO}~\cite{instruments6020019}. 

Given the $\mathcal{O}(10^{-4})$ suppression of the ${}^3\overline{\mathrm{He}}$ multiplicity relative to $\overline{\mathrm{D}}$ in DM annihilation~\cite{DiMauro:2025vxp}, deriving \sigmav{} limits directly from ${}^3\overline{\mathrm{He}}$ null detection analyses is computationally prohibitive. We thus restrict our evaluation to a representative set of DM masses.

\begin{figure}
    \centering
   \includegraphics[width=1\linewidth]{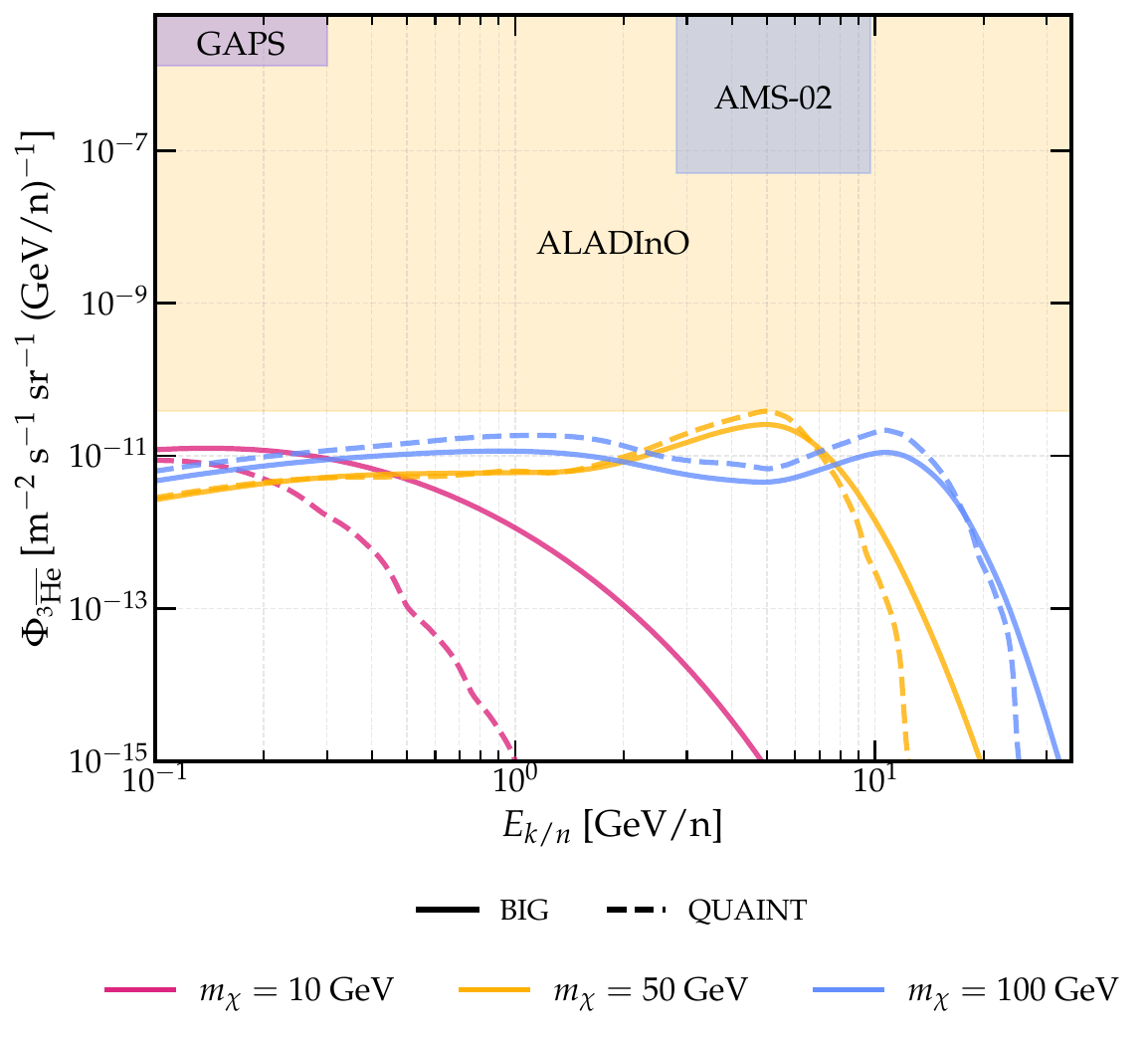}
    \caption{\textbf{Primary fluxes of TOA antihelium 3}, considering $m_\chi=10, 50$, and $100$\,GeV and $\phi_{\rm FF}=750$\,MV, for the \BIG{} and \QUAINT{} models and using the Argonne coalescence approach.  The fluxes are computed considering the 95 $\%$ CL upper limits on \sigmav{} obtained by our antiproton analysis.}
    \label{fig:He3barTOA}
\end{figure}

As shown in Fig.~\ref{fig:He3barTOA}, the predicted TOA flux of primary \hebar\ (for DM DM $\to b\bar{b}$) lies approximately three (four) orders of magnitude below the sensitivity of \textsf{AMS-02} (\textsf{GAPS}). However, depending on the assumed propagation model, the flux could approach the sensitivity of the proposed \textsf{ALADInO} experiment.

An open issue concerns the unpublished claims by the \textsf{AMS-02} collaboration regarding the detection of a dozen \hebar\ events~\cite{Tingcern2016,Miapp2022Dbar,Miapp2022DbarHebar}, which cannot be explained within our primary \hebar\  production framework. Proposed mechanisms invoking an enhancement of \hebar\  production from the decay of weakly-decaying $b$-hadrons~\cite{Winkler:2020ltd,Winkler:2021cmt} appear to be in tension with the recent upper limit on the inclusive branching ratio of $\overline{\Lambda_b^0}$ into \hebar\  reported by the \textsf{LHCb} collaboration~\cite{Moise:2024wqy}, as discussed in Ref.~\cite{DiMauro:2025vxp}. Other approaches that enhance the antinuclei production have been recently discussed in Ref. \cite{dimauro2026enhancedcosmicrayantinucleifluxes}.

\section{Conclusions}

In this work, we have revisited the production and detection prospects of cosmic-ray antinuclei---antiprotons, antideuterons, and antihelium---originating from both standard astrophysical processes and DM annihilation in the Galaxy. Building upon recent theoretical and experimental developments, we have provided updated predictions that incorporate improved modeling of particle production, coalescence mechanisms, and Galactic propagation.
In particular, all the DM fluxes are the maximal ones allowed 
by the \textsf{AMS-02} data on the antiproton flux, fitted together a prediction for the  secondary component.

On the production side, we employed state-of-the-art Monte Carlo simulations, to derive primary antinuclei yields from DM annihilation. We complemented this with updated differential cross-sections for secondary antideuteron production and a systematic comparison of coalescence approaches. In particular, the Argonne Wigner formalism provides a physically motivated benchmark for primary fluxes, while the residual model dependence is found to be at the level of $\mathcal{O}(20\%)$, consistent with recent studies. We have also implemented dedicated \texttt{PYTHIA} tunings, including the recent $\Lambda_b$-informed calibration.

For Galactic transport, we adopted the semi-analytical \usine{} framework and explored multiple propagation scenarios (\BIG{} and \QUAINT{}). These configurations provide comparably good fits to nuclei data but lead to non-negligible differences in the predicted secondary antiproton flux, which in turn impact the inferred DM constraints.

Using the \textsf{AMS-02} antiproton data (2011--2018), we derived updated $95\%$ CL upper limits on the DM annihilation cross-section \sigmav{} over the mass range $m_\chi \sim 5~\mathrm{GeV}$--$10\,\mathrm{TeV}$. The resulting bounds reach $\sigmav \sim \text{few} \times 10^{-26}\,\mathrm{cm}^3\,\mathrm{s}^{-1}$ at $m_\chi \sim 100\,\mathrm{GeV}$, with a clear dependence on the propagation setup: the \BIG{} model yields weaker constraints below $m_\chi \lesssim 20\,\mathrm{GeV}$, while \QUAINT{} allows for larger DM contributions at $m_\chi \gtrsim 300\,\mathrm{GeV}$. 

For the first time, we showed that the inclusion of simulated low-energy antiproton data from \textsf{GAPS}, could improve the limits by more than one order of magnitude for light DM candidates ($m_\chi \lesssim 50\,\mathrm{GeV}$). Alternatively, this same result implies that, for this DM mass range, \textsf{GAPS} can potentially detect a DM signal in the antiproton channel. 

We then translated these constraints into predictions for antideuterons. At sub-GeV/n energies, the maximum allowed primary flux can exceed the secondary background by up to $\sim 2$--$3$ orders of magnitude for light DM candidates ($m_\chi \sim 10~\mathrm{GeV}$), confirming the well-known advantage of this channel. However, detectability strongly depends on the propagation model: within the \BIG{} scenario, \textsf{GAPS} could detect $\mathcal{O}(1)$ primary antideuterons for $m_\chi \lesssim 10~\mathrm{GeV}$, whereas in the \QUAINT{} setup the signal remains below the experimental sensitivity over the full mass range. \textsf{AMS-02} retains sensitivity up to $m_\chi \sim 200~\mathrm{GeV}$. The updated $\Lambda_b$ tuning significantly suppresses the predicted flux for $m_\chi \lesssim 10$--$15~\mathrm{GeV}$, pushing the expected signal below the detection threshold of both \textsf{GAPS} and \textsf{AMS-02} in this region, for the $b\bar b$ annihilation channel.

We also quantified the projected impact of null detections. A non-observation of antideuterons by \textsf{AMS-02} could strengthen current antiproton-derived bounds by up to a factor of $\sim 3$--$5$ for $m_\chi \lesssim 200~\mathrm{GeV}$, while \textsf{GAPS} would only improve constraints in the very low-mass regime and under favorable propagation assumptions. Future experiments such as \textsf{ALADInO} could probe cross-sections close to the thermal relic benchmark up to $m_\chi \sim 1~\mathrm{TeV}$.

For antihelium, we found that the predicted fluxes from DM annihilation are $\sim 3$--$4$ orders of magnitude below the sensitivities of \textsf{AMS-02} and \textsf{GAPS}, although they may approach the reach of next-generation instruments. These results highlight a persistent tension with the tentative \textsf{AMS-02} antihelium candidate events, which cannot be accommodated within standard coalescence-based production models for WIMPs. Nevertheless, enhanced antihelium production can be obtained in models endowed with confining dark sectors, like the SUEP model recently discussed in Ref. \cite{dimauro2026enhancedcosmicrayantinucleifluxes}.

In conclusion, our analysis provides a unified and updated framework for interpreting cosmic-ray antinuclei measurements and emphasizes the complementarity between antiprotons and heavier antinuclei. Actually, several experiments are ongoing (or planned) to both improve the antiproton data and detect $A>1$ antinuclei in cosmic rays. However, the foreseen difficulty to decrease the modelling uncertainties on the secondary antiproton flux~\cite{Maurin:2025gsz} makes heavier antinuclei the only guaranteed route to improve significantly (by up to two orders of magnitude) the DM constraints at low masses.

\begin{acknowledgments} 
M.D.M., F.D., N.F., J.K. and L.S.~acknowledge support from the research grant {\sl TAsP (Theoretical Astroparticle Physics)} funded by Istituto Nazionale di Fisica Nucleare (INFN). M.D.M.~and J.K.~acknowledge support  from the Italian Ministry of University and Research (MUR), PRIN 2022 ``EXSKALIBUR – Euclid-Cross-SKA: Likelihood Inference Building for Universe’s Research'', Grant No. 20222BBYB9, CUP I53D23000610 0006, and from the European Union---Next Generation EU. J.K.~acknowledges support from the Italian Space Agency through the ASI INFN agreement n. 2018-28-HH.0: “Partecipazione italiana al GAPS - General AntiParticle Spectrometer”. N.F.~is supported by the Italian Ministry of University and Research (\textsc{mur}) via the PRIN 2022 Project No. 20228WHTYC – CUP: D53C24003550006 and from the European Union---Next Generation EU.

\end{acknowledgments}

\bibliographystyle{apsrev4-1}
\bibliography{main.bib}

@article{Wiringa:1994wb,
    author = "Wiringa, Robert B. and Stoks, V. G. J. and Schiavilla, R.",
    title = "{An Accurate nucleon-nucleon potential with charge independence breaking}",
    eprint = "nucl-th/9408016",
    archivePrefix = "arXiv",
    reportNumber = "PHY-7742-TH-94, CEBAF-TH-94-19",
    doi = "10.1103/PhysRevC.51.38",
    journal = "Phys. Rev. C",
    volume = "51",
    pages = "38--51",
    year = "1995"
}

@article{DiMauro:2026oto,
  title = {Toward universal coalescence models for antideuteron production},
  author = {Di Mauro, Mattia and Koechler, Jordan and Stefanuto, Lorenzo and Bellini, Francesca and Donato, Fiorenza and Fornengo, Nicolao},
  journal = {Phys. Rev. D},
  volume = {113},
  issue = {12},
  pages = {123011},
  numpages = {18},
  year = {2026},
  month = {Jun},
  publisher = {American Physical Society},
  doi = {10.1103/pytl-mmbf},
  url = {https://link.aps.org/doi/10.1103/pytl-mmbf}
}

@article{PAMELA:2008gwm,
    author = "Adriani, Oscar and others",
    collaboration = "PAMELA",
    title = "{An anomalous positron abundance in cosmic rays with energies 1.5-100 GeV}",
    eprint = "0810.4995",
    archivePrefix = "arXiv",
    primaryClass = "astro-ph",
    doi = "10.1038/nature07942",
    journal = "Nature",
    volume = "458",
    pages = "607--609",
    year = "2009"
}

@article{PAMELA:2017bna,
    author = "Adriani, O. and others",
    collaboration = "PAMELA",
    title = "{Ten years of PAMELA in space}",
    eprint = "1801.10310",
    archivePrefix = "arXiv",
    primaryClass = "astro-ph.HE",
    doi = "10.1393/ncr/i2017-10140-x",
    journal = "Riv. Nuovo Cim.",
    volume = "40",
    number = "10",
    pages = "473--522",
    year = "2017"
}

@article{Osteria:2020lxn,
    author = "Osteria, G.",
    editor = "Krammer, Manfred and Bergauer, Thomas and Dragicevic, Marko and Friedl, Markus and Jeitler, Manfred and Schieck, Jochen and Schwanda, Christoph",
    collaboration = "GAPS",
    title = "{GAPS: A balloon-borne cosmic-ray antimatter experiment}",
    doi = "10.1016/j.nima.2019.05.042",
    journal = "Nucl. Instrum. Meth. A",
    volume = "958",
    pages = "162201",
    year = "2020"
}

@article{Abe:2012tz,
    author = "Abe, K. and others",
    title = "{Search for Antihelium with the BESS-Polar Spectrometer}",
    eprint = "1201.2967",
    archivePrefix = "arXiv",
    primaryClass = "astro-ph.CO",
    doi = "10.1103/PhysRevLett.108.131301",
    journal = "Phys. Rev. Lett.",
    volume = "108",
    pages = "131301",
    year = "2012"
}

@dataset{di_mauro_2026_19099608,
  author       = {Di Mauro, Mattia and
                  Koechler, Jordan and
                  Stefanuto, Lorenzo and
                  Bellini, Francesca and
                  Donato, Fiorenza and
                  Fornengo, Nicolao},
  title        = {Differential cross-sections for secondary
                   antiproton and antideuteron production
                  },
  month        = mar,
  year         = 2026,
  publisher    = {Zenodo},
  doi          = {10.5281/zenodo.19099608},
  url          = {https://doi.org/10.5281/zenodo.19099608},
}

@article{ALICE:2022zuz,
    author = "Acharya, Shreyasi and others",
    collaboration = "ALICE",
    title = "{Measurement of anti-$^{3}$He nuclei absorption in matter and impact on their propagation in the Galaxy}",
    eprint = "2202.01549",
    archivePrefix = "arXiv",
    primaryClass = "nucl-ex",
    reportNumber = "CERN-EP-2022-023",
    doi = "10.1038/s41567-022-01804-8",
    journal = "Nature Phys.",
    volume = "19",
    number = "1",
    pages = "61--71",
    year = "2023"
}

@article{Chardonnet:1997dv,
      author         = "Chardonnet, Pascal and Orloff, Jean and Salati, Pierre",
      title          = "{The Production of antimatter in our galaxy}",
      journal        = "Phys. Lett.",
      volume         = "B409",
      year           = "1997",
      pages          = "313-320",
      doi            = "10.1016/S0370-2693(97)00870-8",
      eprint         = "astro-ph/9705110",
      archivePrefix  = "arXiv",
      primaryClass   = "astro-ph",
      reportNumber   = "ENSLAPP-A-643-97",
      SLACcitation   = "%%CITATION = ASTRO-PH/9705110;%%"
}

@article{DiMauro:2024kml,
    author = "Di Mauro, Mattia and Fornengo, Nicolao and Jueid, Adil and de Austri, Roberto Ruiz and Bellini, Francesca",
    title = "{Nailing Down the Theoretical Uncertainties of D{\textasciimacron} Spectrum Produced from Dark Matter}",
    eprint = "2411.04815",
    archivePrefix = "arXiv",
    primaryClass = "astro-ph.HE",
    reportNumber = "CTPU-PTC-24-31, CERN-TH-2024-164",
    doi = "10.1103/w6n5-vs4d",
    journal = "Phys. Rev. Lett.",
    volume = "135",
    number = "13",
    pages = "131002",
    year = "2025"
}

@article{Orusa:2022pvp,
    author = "Orusa, Luca and Di Mauro, Mattia and Donato, Fiorenza and Korsmeier, Michael",
    title = "{New determination of the production cross section for secondary positrons and electrons in the Galaxy}",
    eprint = "2203.13143",
    archivePrefix = "arXiv",
    primaryClass = "astro-ph.HE",
    doi = "10.1103/PhysRevD.105.123021",
    journal = "Phys. Rev. D",
    volume = "105",
    number = "12",
    pages = "123021",
    year = "2022"
}

@article{Korsmeier:2018gcy,
    author = "Korsmeier, Michael and Donato, Fiorenza and Di Mauro, Mattia",
    title = "{Production cross sections of cosmic antiprotons in the light of new data from the NA61 and LHCb experiments}",
    eprint = "1802.03030",
    archivePrefix = "arXiv",
    primaryClass = "astro-ph.HE",
    reportNumber = "TTK-18-06",
    doi = "10.1103/PhysRevD.97.103019",
    journal = "Phys. Rev. D",
    volume = "97",
    number = "10",
    pages = "103019",
    year = "2018"
}

@unpublished{Oliva:JENAA:2024,
  author       = {Oliva, A.},
  title        = {Latest Results of the {Alpha Magnetic Spectrometer} on the {International Space Station}},
  note         = {Invited talk at the JENAA Workshop @ CERN, 20 Aug 2024},
  organization = {INFN Bologna},
  address      = {CERN, Geneva, Switzerland},
  year         = {2024},
  month        = aug
}

@article{gaps_hailey_cern,
  author       = {Hailey, C.},
  collaboration = "GAPS",
  title        = "{Current Status and Future Prospects of the General Antiparticle Spectrometer (GAPS) Experiment}",
  journal       = "Invited talk at the Workshop https://indico.cern.ch/event/1480110/overview, CERN 22 Jan 2026",
  organization = {CERN},
  address      = {CERN, Geneva, Switzerland},
  year         = {2026},
  month        = jan
}

@article{Korsmeier:2017xzj,
    author = "Korsmeier, Michael and Donato, Fiorenza and Fornengo, Nicolao",
    title = "{Prospects to verify a possible dark matter hint in cosmic antiprotons with antideuterons and antihelium}",
    eprint = "1711.08465",
    archivePrefix = "arXiv",
    primaryClass = "astro-ph.HE",
    reportNumber = "TTK-17-42",
    doi = "10.1103/PhysRevD.97.103011",
    journal = "Phys. Rev. D",
    volume = "97",
    number = "10",
    pages = "103011",
    year = "2018"
}

@article{Strong:2007nh,
    author = "Strong, Andrew W. and Moskalenko, Igor V. and Ptuskin, Vladimir S.",
    title = "{Cosmic-ray propagation and interactions in the Galaxy}",
    eprint = "astro-ph/0701517",
    archivePrefix = "arXiv",
    reportNumber = "SLAC-PUB-12312",
    doi = "10.1146/annurev.nucl.57.090506.123011",
    journal = "Ann. Rev. Nucl. Part. Sci.",
    volume = "57",
    pages = "285--327",
    year = "2007"
}

@article{Kachelriess:2020uoh,
    author = "Kachelrie\ss{}, M. and Ostapchenko, S. and Tjemsland, J.",
    title = "{Revisiting cosmic ray antinuclei fluxes with a new coalescence model}",
    eprint = "2002.10481",
    archivePrefix = "arXiv",
    primaryClass = "hep-ph",
    doi = "10.1088/1475-7516/2020/08/048",
    journal = "JCAP",
    volume = "08",
    pages = "048",
    year = "2020"
}

@article{vonDoetinchem:2015yva,
    author = "von Doetinchem, Philip and others",
    title = "{Status of cosmic-ray antideuteron searches}",
    eprint = "1507.02712",
    archivePrefix = "arXiv",
    primaryClass = "hep-ph",
    doi = "10.22323/1.236.1218",
    journal = "PoS",
    volume = "ICRC2015",
    pages = "1218",
    year = "2016"
}

@article{DeLaTorreLuque:2024htu,
    author = "De La Torre Luque, Pedro and Winkler, Martin Wolfgang and Linden, Tim",
    title = "{Cosmic-ray propagation models elucidate the prospects for antinuclei detection}",
    eprint = "2404.13114",
    archivePrefix = "arXiv",
    primaryClass = "astro-ph.HE",
    doi = "10.1088/1475-7516/2024/10/017",
    journal = "JCAP",
    volume = "10",
    pages = "017",
    year = "2024"
}

@article{ALICE:2025antideuteronRapidity13TeV,
  author  = "{ALICE Collaboration}",
  title   = "{Rapidity dependence of antideuteron coalescence in pp collisions at $\sqrt{s}=13$ TeV with ALICE}",
  journal = "Phys. Lett. B",
  volume  = "860",
  pages   = "139191",
  year    = "2025",
  doi     = "10.1016/j.physletb.2024.139191"
}

@article{ALICE:2019multiplicitypPbLightNuclei,
  author        = "{ALICE Collaboration}",
  title         = "{Multiplicity dependence of light (anti-)nuclei production in p--Pb collisions at $\sqrt{s_{\rm NN}}=5.02$ TeV}",
  journal       = "Phys. Lett. B",
  volume        = "800",
  pages         = "135043",
  year          = "2019",
  doi           = "10.1016/j.physletb.2019.135043",
  eprint        = "1906.03136",
  archivePrefix = "arXiv",
  primaryClass  = "nucl-ex"
}

@Article{Serksnyte:2022onw,
  author        = {\v{S}erk\v{s}nyt\.{e}, Laura and others},
  journal       = {Phys. Rev. D},
  title         = {{Reevaluation of the cosmic antideuteron flux from cosmic-ray interactions and from exotic sources}},
  year          = {2022},
  number        = {8},
  pages         = {083021},
  volume        = {105},
  archiveprefix = {arXiv},
  doi           = {10.1103/PhysRevD.105.083021},
  eprint        = {2201.00925},
  primaryclass  = {astro-ph.HE},
}

@Article{Calore:2022stf,
  author        = {Calore, Francesca and Cirelli, Marco and Derome, Laurent and Genolini, Yoann and Maurin, David and Salati, Pierre and Serpico, Pasquale Dario},
  journal       = {SciPost Phys.},
  title         = {{AMS-02 antiprotons and dark matter: Trimmed hints and robust bounds}},
  year          = {2022},
  number        = {5},
  pages         = {163},
  volume        = {12},
  archiveprefix = {arXiv},
  doi           = {10.21468/SciPostPhys.12.5.163},
  eprint        = {2202.03076},
  primaryclass  = {hep-ph},
  reportnumber  = {LAPTH-003/22},
}

@Article{AMS:2021nhj,
  author        = {Aguilar, M. and others},
  journal       = {Phys. Rept.},
  title         = {{The Alpha Magnetic Spectrometer (AMS) on the international space station: Part II \textemdash{} Results from the first seven years}},
  year          = {2021},
  pages         = {1--116},
  volume        = {894},
  collaboration = {AMS},
  doi           = {10.1016/j.physrep.2020.09.003},
}

@Misc{Tingcern2016,
  author       = {S. Ting},
  howpublished = {CERN},
  title        = {The First Five Years of the Alpha Magnetic Spectrometer on the International Space Station: Unlocking the Secrets of the Cosmos},
  year         = {2016},
  url          = {https://indico.cern.ch/event/592392/attachments/1381599/2110332/AMS-CERN-Dec-2016.pdf},
}

@Misc{Miapp2022DbarHebar,
  author       = {P. Zuccon},
  howpublished = {MIAPP},
  title        = {AMS-02 results \& upgrade},
  year         = {2022},
  url          = {https://indico.ph.tum.de/event/6990/contributions/4988/attachments/3947/4992/Zuccon_miapp.pdf},
}

@Misc{Miapp2022Dbar,
  author       = {Senquan Lu},
  howpublished = {MIAPP},
  title        = {Cosmic Ray Antideuteron Search with Alpha Magnetic Spectrometer (AMS)},
  year         = {2022},
  url          = {https://indico.ph.tum.de/event/6990/contributions/4970/attachments/3924/4946/Lu_Slides.pdf},
}

@Article{CREMA:2016idx,
  author        = {Pohl, Randolf and others},
  journal       = {Science},
  title         = {{Laser spectroscopy of muonic deuterium}},
  year          = {2016},
  number        = {6300},
  pages         = {669--673},
  volume        = {353},
  collaboration = {CREMA},
  doi           = {10.1126/science.aaf2468},
}

@Article{PhysRevLett.132.131001,
  author        = {Sakai, K. and others},
  journal       = {Phys. Rev. Lett.},
  title         = {Search for Antideuterons of Cosmic Origin Using the BESS-Polar II Magnetic-Rigidity Spectrometer},
  year          = {2024},
  month         = {Mar},
  pages         = {131001},
  volume        = {132},
  collaboration = {BESS Collaboration},
  doi           = {10.1103/PhysRevLett.132.131001},
  issue         = {13},
  numpages      = {6},
  publisher     = {American Physical Society},
  url           = {https://link.aps.org/doi/10.1103/PhysRevLett.132.131001},
}

@Article{Ibarra:2012cc,
  author        = {Ibarra, Alejandro and Wild, Sebastian},
  journal       = {JCAP},
  title         = {{Prospects of antideuteron detection from dark matter annihilations or decays at AMS-02 and GAPS}},
  year          = {2013},
  pages         = {021},
  volume        = {02},
  archiveprefix = {arXiv},
  doi           = {10.1088/1475-7516/2013/02/021},
  eprint        = {1209.5539},
  primaryclass  = {hep-ph},
}

@Article{Jueid:2022qjg,
  author        = {Jueid, Adil and Kip, Jochem and de Austri, Roberto Ruiz and Skands, Peter},
  journal       = {JCAP},
  title         = {{Impact of QCD uncertainties on antiproton spectra from dark-matter annihilation}},
  year          = {2023},
  pages         = {068},
  volume        = {04},
  archiveprefix = {arXiv},
  doi           = {10.1088/1475-7516/2023/04/068},
  eprint        = {2202.11546},
  primaryclass  = {hep-ph},
  reportnumber  = {CTPU-PTC-22-22, KIAS-Q22004},
}

@Article{HFLAV:2019otj,
  author        = {Amhis, Yasmine Sara and others},
  journal       = {Eur. Phys. J. C},
  title         = {{Averages of b-hadron, c-hadron, and $\tau $-lepton properties as of 2018}},
  year          = {2021},
  number        = {3},
  pages         = {226},
  volume        = {81},
  archiveprefix = {arXiv},
  collaboration = {HFLAV},
  doi           = {10.1140/epjc/s10052-020-8156-7},
  eprint        = {1909.12524},
  primaryclass  = {hep-ex},
}

@Article{Moise:2024wqy,
  author  = {Moise, Razvan-Daniel},
  journal = {PoS},
  title   = {{Antihelium production in $\overline{\it \Lambda}_b^0$ decays}},
  year    = {2025},
  pages   = {676},
  volume  = {ICHEP2024},
  doi     = {10.22323/1.476.0676},
}

@Article{Amoroso:2018qga,
  author        = {Amoroso, Simone and Caron, Sascha and Jueid, Adil and Ruiz de Austri, Roberto and Skands, Peter},
  journal       = {JCAP},
  title         = {{Estimating QCD uncertainties in Monte Carlo event generators for gamma-ray dark matter searches}},
  year          = {2019},
  pages         = {007},
  volume        = {05},
  archiveprefix = {arXiv},
  doi           = {10.1088/1475-7516/2019/05/007},
  eprint        = {1812.07424},
  primaryclass  = {hep-ph},
}

@Article{Jueid:2023vrb,
  author        = {Jueid, Adil and Kip, Jochem and de Austri, Roberto Ruiz and Skands, Peter},
  journal       = {JHEP},
  title         = {{The Strong Force meets the Dark Sector: a robust estimate of QCD uncertainties for anti-matter dark matter searches}},
  year          = {2024},
  pages         = {119},
  volume        = {02},
  archiveprefix = {arXiv},
  doi           = {10.1007/JHEP02(2024)119},
  eprint        = {2303.11363},
  primaryclass  = {hep-ph},
  reportnumber  = {CTPU-PTC-23-08},
}

@Article{Winkler:2021cmt,
  author        = {Winkler, Martin Wolfgang and Linden, Tim},
  title         = {{Response to Comment on ''Dark Matter Annihilation Can Produce a Detectable Antihelium Flux through $\bar{\Lambda}_b$ Decays''}},
  year          = {2021},
  month         = {5},
  archiveprefix = {arXiv},
  eprint        = {2106.00053},
  primaryclass  = {hep-ph},
}

@Article{DiMauro:2025vxp,
  author        = {Di Mauro, Mattia and Jueid, Adil and Koechler, Jordan and de Austri, Roberto Ruiz},
  journal       = {Phys. Rev. D},
  title         = {{Robust determination of antinuclei production from dark matter via weakly decaying beauty hadrons}},
  year          = {2025},
  number        = {8},
  pages         = {083017},
  volume        = {112},
  archiveprefix = {arXiv},
  doi           = {10.1103/s6cm-45b4},
  eprint        = {2504.07172},
  primaryclass  = {hep-ph},
}

@Article{ALEPH:2006qoi,
  author        = {Schael, S. and others},
  journal       = {Phys. Lett. B},
  title         = {{Deuteron and anti-deuteron production in e+ e- collisions at the Z resonance}},
  year          = {2006},
  pages         = {192--201},
  volume        = {639},
  archiveprefix = {arXiv},
  collaboration = {ALEPH},
  doi           = {10.1016/j.physletb.2006.06.043},
  eprint        = {hep-ex/0604023},
  reportnumber  = {CERN-PH-EP-2006-009},
}

@Article{Aramaki:2015laa,
  author        = {Aramaki, T. and Hailey, C. J. and Boggs, S. E. and von Doetinchem, P. and Fuke, H. and Mognet, S. I. and Ong, R. A. and Perez, K. and Zweerink, J.},
  journal       = {Astropart. Phys.},
  title         = {{Antideuteron Sensitivity for the GAPS Experiment}},
  year          = {2016},
  pages         = {6--13},
  volume        = {74},
  archiveprefix = {arXiv},
  collaboration = {GAPS},
  doi           = {10.1016/j.astropartphys.2015.09.001},
  eprint        = {1506.02513},
  primaryclass  = {astro-ph.HE},
}

@article{DiMauro:2023jgg,
    author = "Di Mauro, Mattia and Korsmeier, Michael and Cuoco, Alessandro",
    title = "{Data-driven constraints on cosmic-ray diffusion: Probing self-generated turbulence in the Milky~Way}",
    eprint = "2311.17150",
    archivePrefix = "arXiv",
    primaryClass = "astro-ph.HE",
    doi = "10.1103/PhysRevD.109.123003",
    journal = "Phys. Rev. D",
    volume = "109",
    number = "12",
    pages = "123003",
    year = "2024"
}

@Article{Fornengo:2013osa,
  author        = {Fornengo, N. and Maccione, L. and Vittino, A.},
  journal       = {JCAP},
  title         = {{Dark matter searches with cosmic antideuterons: status and perspectives}},
  year          = {2013},
  pages         = {031},
  volume        = {09},
  archiveprefix = {arXiv},
  doi           = {10.1088/1475-7516/2013/09/031},
  eprint        = {1306.4171},
  primaryclass  = {hep-ph},
  reportnumber  = {LMU-ASC-41-13, MPP-2013-158},
}

@Article{vonDoetinchem:2020vbj,
  author        = {von Doetinchem, P. and others},
  journal       = {JCAP},
  title         = {{Cosmic-ray antinuclei as messengers of new physics: status and outlook for the new decade}},
  year          = {2020},
  pages         = {035},
  volume        = {08},
  archiveprefix = {arXiv},
  doi           = {10.1088/1475-7516/2020/08/035},
  eprint        = {2002.04163},
  primaryclass  = {astro-ph.HE},
}

@Article{Winkler:2020ltd,
  author        = {Winkler, Martin Wolfgang and Linden, Tim},
  journal       = {Phys. Rev. Lett.},
  title         = {{Dark Matter Annihilation Can Produce a Detectable Antihelium Flux through $\bar{\Lambda}_b$ Decays}},
  year          = {2021},
  number        = {10},
  pages         = {101101},
  volume        = {126},
  archiveprefix = {arXiv},
  doi           = {10.1103/PhysRevLett.126.101101},
  eprint        = {2006.16251},
  primaryclass  = {hep-ph},
}

@Article{Scheibl:1998tk,
  author        = {Scheibl, Rudiger and Heinz, Ulrich W.},
  journal       = {Phys. Rev.},
  title         = {{Coalescence and flow in ultrarelativistic heavy ion collisions}},
  year          = {1999},
  pages         = {1585-1602},
  volume        = {C59},
  archiveprefix = {arXiv},
  doi           = {10.1103/PhysRevC.59.1585},
  eprint        = {nucl-th/9809092},
  primaryclass  = {nucl-th},
  slaccitation  = {%%CITATION = NUCL-TH/9809092;%%},
}

@Article{Blum:2017qnn,
  author        = {Blum, Kfir and Ng, Kenny Chun Yu and Sato, Ryosuke and Takimoto, Masahiro},
  journal       = {Phys. Rev. D},
  title         = {{Cosmic rays, antihelium, and an old navy spotlight}},
  year          = {2017},
  number        = {10},
  pages         = {103021},
  volume        = {96},
  archiveprefix = {arXiv},
  doi           = {10.1103/PhysRevD.96.103021},
  eprint        = {1704.05431},
  primaryclass  = {astro-ph.HE},
}

@Article{Bellini:2020cbj,
  author        = {Bellini, Francesca and Blum, Kfir and Kalweit, Alexander Phillip and Puccio, Maximiliano},
  journal       = {Phys. Rev. C},
  title         = {{Examination of coalescence as the origin of nuclei in hadronic collisions}},
  year          = {2021},
  number        = {1},
  pages         = {014907},
  volume        = {103},
  archiveprefix = {arXiv},
  doi           = {10.1103/PhysRevC.103.014907},
  eprint        = {2007.01750},
  primaryclass  = {nucl-th},
}

@Article{Bellini:2018epz,
  author        = {Bellini, Francesca and Kalweit, Alexander Philipp},
  journal       = {Phys. Rev. C},
  title         = {{Testing production scenarios for (anti-)(hyper-)nuclei and exotica at energies available at the CERN Large Hadron Collider}},
  year          = {2019},
  number        = {5},
  pages         = {054905},
  volume        = {99},
  archiveprefix = {arXiv},
  doi           = {10.1103/PhysRevC.99.054905},
  eprint        = {1807.05894},
  primaryclass  = {hep-ph},
}

@Article{Kachelriess:2019taq,
  author        = {Kachelrie\ss{}, M. and Ostapchenko, S. and Tjemsland, J.},
  journal       = {Eur. Phys. J. A},
  title         = {{Alternative coalescence model for deuteron, tritium, helium-3 and their antinuclei}},
  year          = {2020},
  number        = {1},
  pages         = {4},
  volume        = {56},
  archiveprefix = {arXiv},
  doi           = {10.1140/epja/s10050-019-00007-9},
  eprint        = {1905.01192},
  primaryclass  = {hep-ph},
}

@INPROCEEDINGS{2008ICRC....4..765C,
       author = {{Choutko}, V. and {Giovacchini}, F.},
        title = "{Cosmic Rays Antideuteron Sensitivity for AMS-02 Experiment}",
    booktitle = {International Cosmic Ray Conference},
         year = 2008,
       series = {International Cosmic Ray Conference},
       volume = {4},
        month = jan,
        pages = {765-768},
       adsurl = {https://ui.adsabs.harvard.edu/abs/2008ICRC....4..765C}
}

@Article{Carlson:2014ssa,
  author        = {Carlson, Eric and Coogan, Adam and Linden, Tim and Profumo, Stefano and Ibarra, Alejandro and Wild, Sebastian},
  journal       = {Phys. Rev. D},
  title         = {{Antihelium from Dark Matter}},
  year          = {2014},
  number        = {7},
  pages         = {076005},
  volume        = {89},
  archiveprefix = {arXiv},
  doi           = {10.1103/PhysRevD.89.076005},
  eprint        = {1401.2461},
  primaryclass  = {hep-ph},
}

@article{Bierlich:2022pfr,
    author = "Bierlich, Christian and others",
    title = "{A comprehensive guide to the physics and usage of PYTHIA 8.3}",
    eprint = "2203.11601",
    archivePrefix = "arXiv",
    primaryClass = "hep-ph",
    reportNumber = "LU-TP 22-16, MCNET-22-04, FERMILAB-PUB-22-227-SCD",
    doi = "10.21468/SciPostPhysCodeb.8",
    journal = "SciPost Phys. Codeb.",
    volume = "2022",
    pages = "8",
    year = "2022"
}

@Article{Donato:1999gy,
  author        = {Donato, Fiorenza and Fornengo, Nicolao and Salati, Pierre},
  journal       = {Physical Review D},
  title         = {Antideuterons as a Signature of Supersymmetric Dark Matter},
  year          = {2000},
  pages         = {043003},
  volume        = {62},
  archiveprefix = {arXiv},
  doi           = {10.1103/PhysRevD.62.043003},
  eprint        = {hep-ph/9904481},
  primaryclass  = {hep-ph},
  reportnumber  = {IFIC-99-9, FTUV-99-9, LAPTH-722-99},
  slaccitation  = {%%CITATION = HEP-PH/9904481;%%},
}

@Article{Cirelli:2014qia,
  author        = {Cirelli, Marco and Fornengo, Nicolao and Taoso, Marco and Vittino, Andrea},
  journal       = {Journal of High Energy Physics},
  title         = {Anti-helium from dark matter annihilations},
  year          = {2014},
  number        = {08},
  pages         = {009},
  volume        = {2014},
  archiveprefix = {arXiv},
  doi           = {10.1007/JHEP08(2014)009},
  eprint        = {1401.4017},
  primaryclass  = {hep-ph},
  reportnumber  = {SACLAY-T14-003},
}

@Article{ALICE:2020foi,
  author        = {Acharya, S. and others},
  journal       = {Eur. Phys. J. C},
  title         = {{(Anti-)deuteron production in pp collisions at $\sqrt{s}=13 \ \text {TeV}$}},
  year          = {2020},
  number        = {9},
  pages         = {889},
  volume        = {80},
  archiveprefix = {arXiv},
  collaboration = {ALICE},
  doi           = {10.1140/epjc/s10052-020-8256-4},
  eprint        = {2003.03184},
  primaryclass  = {nucl-ex},
  reportnumber  = {CERN-EP-2020-025},
}

@Article{Genolini:2021doh,
  author        = {G\'enolini, Yoann and Boudaud, Mathieu and Cirelli, Marco and Derome, Laurent and Lavalle, Julien and Maurin, David and Salati, Pierre and Weinrich, Nathanael},
  journal       = {Phys. Rev. D},
  title         = {{New minimal, median, and maximal propagation models for dark matter searches with Galactic cosmic rays}},
  year          = {2021},
  month         = {3},
  number        = {8},
  pages         = {083005},
  volume        = {104},
  archiveprefix = {arXiv},
  doi           = {10.1103/PhysRevD.104.083005},
  eprint        = {2103.04108},
  primaryclass  = {astro-ph.HE},
  reportnumber  = {LAPTH-011/21, LUPM:21-002},
}

@Article{Balan:2023lwg,
  author        = {Balan, Sowmiya and Kahlhoefer, Felix and Korsmeier, Michael and Manconi, Silvia and Nippel, Kathrin},
  journal       = {JCAP},
  title         = {{Fast and accurate AMS-02 antiproton likelihoods for global dark matter fits}},
  year          = {2023},
  month         = {3},
  pages         = {052},
  volume        = {08},
  archiveprefix = {arXiv},
  doi           = {10.1088/1475-7516/2023/08/052},
  eprint        = {2303.07362},
  primaryclass  = {hep-ph},
  reportnumber  = {TTK-23-02, TTP-23-008, LAPTH-008/23},
}

@Article{Maurin:2001sj,
  author        = {Maurin, D. and Donato, F. and Taillet, R. and Salati, P.},
  journal       = {Astrophys. J.},
  title         = {{Cosmic rays below z=30 in a diffusion model: new constraints on propagation parameters}},
  year          = {2001},
  pages         = {585--596},
  volume        = {555},
  archiveprefix = {arXiv},
  doi           = {10.1086/321496},
  eprint        = {astro-ph/0101231},
  reportnumber  = {PREPRINT-LAPTH-817-00},
}

@Article{Adriani:2012paa,
  author  = {Adriani, O. and others},
  journal = {JETP Lett.},
  title   = {{Measurement of the flux of primary cosmic ray antiprotons with energies of 60-MeV to 350-GeV in the PAMELA experiment}},
  year    = {2013},
  pages   = {621--627},
  volume  = {96},
  doi     = {10.1134/S002136401222002X},
}

@article{Mahlein:2023fmx,
    author = "Mahlein, Maximilian and Barioglio, Luca and Bellini, Francesca and Fabbietti, Laura and Pinto, Chiara and Singh, Bhawani and Tripathy, Sushanta",
    title = "{A realistic coalescence model for deuteron production}",
    eprint = "2302.12696",
    archivePrefix = "arXiv",
    primaryClass = "hep-ex",
    doi = "10.1140/epjc/s10052-023-11972-3",
    journal = "Eur. Phys. J. C",
    volume = "83",
    number = "9",
    pages = "804",
    year = "2023"
}

@Article{ALICE:2017xrp,
  author        = {Acharya, S. and others},
  journal       = {Phys. Rev. C},
  title         = {Production of deuterons, tritons, $^{3}\mathrm{He}$ nuclei, and their antinuclei in $pp$ collisions at $\sqrt{s}=0.9$, 2.76, and 7 TeV},
  year          = {2018},
  month         = {Feb},
  number        = {2},
  pages         = {024615},
  volume        = {97},
  archiveprefix = {arXiv},
  collaboration = {ALICE Collaboration},
  doi           = {10.1103/PhysRevC.97.024615},
  eprint        = {1709.08522},
  issue         = {2},
  numpages      = {17},
  primaryclass  = {nucl-ex},
  publisher     = {American Physical Society},
  reportnumber  = {CERN-EP-2017-255},
  url           = {https://link.aps.org/doi/10.1103/PhysRevC.97.024615},
}

@Article{Horst:2023oti,
  author        = {Horst, Maximilian and Barioglio, Luca and Bellini, Francesca and Fabbietti, Laura and Pinto, Chiara and Singh, Bhawani and Tripathy, Sushanta},
  journal       = {Eur. Phys. J. C},
  title         = {{Novel parameter-free coalescence model for deuteron production}},
  year          = {2023},
  month         = {2},
  number        = {9},
  pages         = {804},
  volume        = {83},
  archiveprefix = {arXiv},
  doi           = {10.1140/epjc/s10052-023-11972-3},
  eprint        = {2302.12696},
  primaryclass  = {hep-ex},
}

@Article{PhysRevLett.117.091103,
  author        = {Aguilar, M. and Ali Cavasonza, L. and Alpat, B. and Ambrosi, G. and others},
  journal       = {Phys. Rev. Lett.},
  title         = {Antiproton Flux, Antiproton-to-Proton Flux Ratio, and Properties of Elementary Particle Fluxes in Primary Cosmic Rays Measured with the Alpha Magnetic Spectrometer on the International Space Station},
  year          = {2016},
  month         = {Aug},
  pages         = {091103},
  volume        = {117},
  collaboration = {AMS Collaboration},
  doi           = {10.1103/PhysRevLett.117.091103},
  issue         = {9},
  numpages      = {10},
  publisher     = {American Physical Society},
  url           = {https://link.aps.org/doi/10.1103/PhysRevLett.117.091103},
}

@Article{Weinrich:2020ftb,
  author        = {Weinrich, N. and Boudaud, M. and Derome, L. and Genolini, Y. and Lavalle, J. and Maurin, D. and Salati, P. and Serpico, P. and Weymann-Despres, G.},
  journal       = {Astron. Astrophys.},
  title         = {{Galactic halo size in the light of recent AMS-02 data}},
  year          = {2020},
  pages         = {A74},
  volume        = {639},
  archiveprefix = {arXiv},
  doi           = {10.1051/0004-6361/202038064},
  eprint        = {2004.00441},
  primaryclass  = {astro-ph.HE},
  reportnumber  = {LAPTH-009/20, LUPM:20-019},
}

@Article{Weinrich:2020cmw,
  author        = {Weinrich, Nathanael and G\'enolini, Yoann and Boudaud, Mathieu and Derome, Laurent and Maurin, David},
  journal       = {Astron. Astrophys.},
  title         = {{Combined analysis of AMS-02 (Li,Be,B)/C, N/O, 3He, and 4He data}},
  year          = {2020},
  pages         = {A131},
  volume        = {639},
  archiveprefix = {arXiv},
  doi           = {10.1051/0004-6361/202037875},
  eprint        = {2002.11406},
  primaryclass  = {astro-ph.HE},
}

@Article{Donato:2003xg,
  author        = {Donato, F. and Fornengo, Nicolao and Maurin, D. and Salati, P.},
  journal       = {Phys. Rev. D},
  title         = {{Antiprotons in cosmic rays from neutralino annihilation}},
  year          = {2004},
  pages         = {063501},
  volume        = {69},
  archiveprefix = {arXiv},
  doi           = {10.1103/PhysRevD.69.063501},
  eprint        = {astro-ph/0306207},
  reportnumber  = {DFTT-32-02, KIAS-P02069},
}

@Article{Boudaud:2019efq,
  author        = {Boudaud, Mathieu and G\'enolini, Yoann and Derome, Laurent and Lavalle, Julien and Maurin, David and Salati, Pierre and Serpico, Pasquale D.},
  journal       = {Phys. Rev. Res.},
  title         = {{AMS-02 antiprotons' consistency with a secondary astrophysical origin}},
  year          = {2020},
  number        = {2},
  pages         = {023022},
  volume        = {2},
  archiveprefix = {arXiv},
  doi           = {10.1103/PhysRevResearch.2.023022},
  eprint        = {1906.07119},
  primaryclass  = {astro-ph.HE},
}

@article{Cholis:2020tpi,
    author = "Cholis, Ilias and Hooper, Dan and Linden, Tim",
    title = "{Constraining the charge-sign and rigidity-dependence of solar modulation}",
    eprint = "2007.00669",
    archivePrefix = "arXiv",
    primaryClass = "astro-ph.HE",
    reportNumber = "FERMILAB-PUB-20-271-AE-T",
    doi = "10.1088/1475-7516/2022/10/051",
    journal = "JCAP",
    volume = "10",
    pages = "051",
    year = "2022"
}

@article{AMS:2016oqu,
    author = "Aguilar, M. and others",
    collaboration = "AMS",
    title = "{Antiproton Flux, Antiproton-to-Proton Flux Ratio, and Properties of Elementary Particle Fluxes in Primary Cosmic Rays Measured with the Alpha Magnetic Spectrometer on the International Space Station}",
    doi = "10.1103/PhysRevLett.117.091103",
    journal = "Phys. Rev. Lett.",
    volume = "117",
    number = "9",
    pages = "091103",
    year = "2016"
}

@article{Adriani:2008zq,
    author = "Adriani, O. and others",
    title = "{A new measurement of the antiproton-to-proton flux ratio up to 100 GeV in the cosmic radiation}",
    eprint = "0810.4994",
    archivePrefix = "arXiv",
    primaryClass = "astro-ph",
    doi = "10.1103/PhysRevLett.102.051101",
    journal = "Phys. Rev. Lett.",
    volume = "102",
    pages = "051101",
    year = "2009"
}

@article{GAPS:2022ncd,
    author = "Rogers, F. and others",
    collaboration = "GAPS",
    title = "{Sensitivity of the GAPS experiment to low-energy cosmic-ray antiprotons}",
    eprint = "2206.12991",
    archivePrefix = "arXiv",
    primaryClass = "astro-ph.HE",
    reportNumber = "FERMILAB-PUB-23-148-PPD",
    doi = "10.1016/j.astropartphys.2022.102791",
    journal = "Astropart. Phys.",
    volume = "145",
    pages = "102791",
    year = "2023"
}

@article{Aramaki_2016,
	doi          = {10.1016/j.astropartphys.2015.09.001},
	year         = 2016,
	month        = {feb},
	publisher    = {Elsevier {BV}},
	volume       = 74,
	pages        = {6--13},
	author       = {T. Aramaki and C.J. Hailey and S.E. Boggs and P. von Doetinchem and H. Fuke and S.I. Mognet and R.A. Ong and K. Perez and others},
	title        = "{Antideuteron sensitivity for the {GAPS} experiment}",
	journal      = {\ap}
}

@article{Arina:2023eic,
    author = "Arina, Chiara and Di Mauro, Mattia and Fornengo, Nicolao and Heisig, Jan and Jueid, Adil and de Austri, Roberto Ruiz",
    title = "{CosmiXs: cosmic messenger spectra for indirect dark matter searches}",
    eprint = "2312.01153",
    archivePrefix = "arXiv",
    primaryClass = "astro-ph.HE",
    reportNumber = "TTK-23-32, CTPU-PTC-23-36",
    doi = "10.1088/1475-7516/2024/03/035",
    journal = "JCAP",
    volume = "03",
    pages = "035",
    year = "2024"
}

@inproceedings{Arina:2025ner,
    author = "Arina, Chiara and Di Mauro, Mattia and Fornengo, Nicolao and Heisig, Jan and Jueid, Adil and de Austri, Roberto Ruiz",
    title = "{CosmiXs: Improved spectra for dark matter indirect detection}",
    eprint = "2501.13281",
    archivePrefix = "arXiv",
    primaryClass = "hep-ph",
    reportNumber = "CTPU-PTC-25-02, IRMP-CP3-25-03, TTK-25-04",
    month = "1",
    year = "2025"
}

@article{Maurin:2018rmm,
    author = "Maurin, David",
    title = "{USINE: semi-analytical models for Galactic cosmic-ray propagation}",
    eprint = "1807.02968",
    archivePrefix = "arXiv",
    primaryClass = "astro-ph.IM",
    doi = "10.1016/j.cpc.2019.106942",
    journal = "Comput. Phys. Commun.",
    volume = "247",
    pages = "106942",
    year = "2020"
}

@article{Donato:2008yx,
    author = "Donato, Fiorenza and Fornengo, Nicolao and Maurin, David",
    title = "{Antideuteron fluxes from dark matter annihilation in diffusion models}",
    eprint = "0803.2640",
    archivePrefix = "arXiv",
    primaryClass = "hep-ph",
    reportNumber = "DFTT-06-2007",
    doi = "10.1103/PhysRevD.78.043506",
    journal = "Phys. Rev. D",
    volume = "78",
    pages = "043506",
    year = "2008"
}

@BOOK{1990acr..book.....B,
       author = {{Berezinskii}, V.~S. and {Bulanov}, S.~V. and {Dogiel}, V.~A. and {Ptuskin}, V.~S.},
        title = "{Astrophysics of cosmic rays}",
         year = 1990,
       adsurl = {https://ui.adsabs.harvard.edu/abs/1990acr..book.....B}
}

@Book{2002cra..book.....S,
  Title                    = {{Cosmic Ray Astrophysics}},
  Author                   = {{Schlickeiser}, R.},
  Year                     = {2002},
  publisher                = {Springer},
  Adsurl                   = {http://adsabs.harvard.edu/abs/2002cra..book.....S},
  Booktitle                = {Cosmic ray astrophysics / Reinhard Schlickeiser, Astronomy and Astrophysics Library; Physics and Astronomy Online Library.~Berlin: Springer.~ISBN 3-540-66465-3, 2002, XV + 519 pp.}
}

@ARTICLE{2019PhRvD..99l3028G,
       author = {{G{\'e}nolini}, Y. and {Boudaud}, M. and {Batista}, P. -I. and {Caroff}, S. and {Derome}, L. and {Lavalle}, J. and {Marcowith}, A. and {Maurin}, D. and {Poireau}, V. and {Poulin}, V. and {Rosier}, S. and {Salati}, P. and {Serpico}, P.~D. and {Vecchi}, M.},
        title = "{Cosmic-ray transport from AMS-02 boron to carbon ratio data: Benchmark models and interpretation}",
      journal = {\prd},
         year = 2019,
        month = jun,
       volume = {99},
       number = {12},
          eid = {123028},
        pages = {123028},
          doi = {10.1103/PhysRevD.99.123028},
archivePrefix = {arXiv},
       eprint = {1904.08917},
 primaryClass = {astro-ph.HE},
       adsurl = {https://ui.adsabs.harvard.edu/abs/2019PhRvD..99l3028G}
}

@article{Maurin:2025gsz,
    author = "Maurin, D. and others",
    title = "{Precision cross-sections for advancing cosmic-ray physics and other applications: A comprehensive programme for the next decade}",
    eprint = "2503.16173",
    archivePrefix = "arXiv",
    primaryClass = "astro-ph.HE",
    doi = "10.1016/j.physrep.2025.11.002",
    journal = "Phys. Rept.",
    volume = "1161",
    pages = "1--81",
    year = "2026"
}

@ARTICLE{2020CoPhC.24706942M,
       author = {{Maurin}, David},
        title = "{USINE: Semi-analytical models for Galactic cosmic-ray propagation}",
      journal = {Computer Physics Communications},
         year = 2020,
        month = feb,
       volume = {247},
          eid = {106942},
        pages = {106942},
          doi = {10.1016/j.cpc.2019.106942},
archivePrefix = {arXiv},
       eprint = {1807.02968},
 primaryClass = {astro-ph.IM},
       adsurl = {https://ui.adsabs.harvard.edu/abs/2020CoPhC.24706942M}
}

@article{Navarro:1995iw,
    author = "Navarro, Julio F. and Frenk, Carlos S. and White, Simon D. M.",
    title = "{The Structure of cold dark matter halos}",
    eprint = "astro-ph/9508025",
    archivePrefix = "arXiv",
    doi = "10.1086/177173",
    journal = "Astrophys. J.",
    volume = "462",
    pages = "563--575",
    year = "1996"
}

@article{McMillan_2016,
   title={The mass distribution and gravitational potential of the Milky Way},
   volume={465},
   ISSN={1365-2966},
   url={http://dx.doi.org/10.1093/mnras/stw2759},
   DOI={10.1093/mnras/stw2759},
   number={1},
   journal={Monthly Notices of the Royal Astronomical Society},
   publisher={Oxford University Press (OUP)},
   author={McMillan, Paul J.},
   year={2016},
   month={Oct},
   pages={76–94}
}

@ARTICLE{2021RPPh...84j4901D,
       author = {{de Salas}, Pablo F. and {Widmark}, A.},
        title = "{Dark matter local density determination: recent observations and future prospects}",
      journal = {Reports on Progress in Physics},
         year = 2021,
        month = oct,
       volume = {84},
       number = {10},
          eid = {104901},
        pages = {104901},
          doi = {10.1088/1361-6633/ac24e7},
archivePrefix = {arXiv},
       eprint = {2012.11477},
 primaryClass = {astro-ph.GA},
       adsurl = {https://ui.adsabs.harvard.edu/abs/2021RPPh...84j4901D}
}

@ARTICLE{2019AandA...625L..10G,
       author = {{Gravity Collaboration} and {Abuter}, R. and others },
        title = "{A geometric distance measurement to the Galactic center black hole with 0.3\% uncertainty}",
      journal = "Astron. Astrophys.",
         year = 2019,
        month = may,
       volume = {625},
          eid = {L10},
        pages = {L10},
          doi = {10.1051/0004-6361/201935656},
archivePrefix = {arXiv},
       eprint = {1904.05721},
 primaryClass = {astro-ph.GA},
       adsurl = {https://ui.adsabs.harvard.edu/abs/2019A&A...625L..10G}
}

@Article{GleesonEtAl1968a,
  author    = {{Gleeson}, L.~J. and {Axford}, W.~I.},
  title     = {{Solar Modulation of Galactic Cosmic Rays}},
  journal   = {\apj},
  year      = {1968},
  volume    = {154},
  pages     = {1011},
  month     = dec,
  adsurl    = {http://adsabs.harvard.edu/abs/1968ApJ...154.1011G},
  doi       = {10.1086/149822},
}

@Article{Fisk1971,
  author    = {{Fisk}, L.~A.},
  title     = {{Solar modulation of galactic cosmic rays, 2}},
  journal   = {J. Geophys. Res.},
  year      = {1971},
  volume    = {76},
  pages     = {221},
  adsurl    = {http://adsabs.harvard.edu/abs/1971JGR....76..221F},
  doi       = {10.1029/JA076i001p00221},
}

@ARTICLE{GhelfiEtAl2017,
       author = {{Ghelfi}, A. and {Maurin}, D. and {Cheminet}, A. and {Derome}, L. and
         {Hubert}, G. and {Melot}, F.},
        title = "{Neutron monitors and muon detectors for solar modulation studies: 2. {\ensuremath{\phi}} time series}",
      journal = {Advances in Space Research},
         year = 2017,
        month = aug,
       volume = {60},
       number = {4},
        pages = {833-847},
          doi = {10.1016/j.asr.2016.06.027},
archivePrefix = {arXiv},
       eprint = {1607.01976},
 primaryClass = {astro-ph.HE},
       adsurl = {https://ui.adsabs.harvard.edu/abs/2017AdSpR..60..833G}
}

@Article{MaurinEtAl2014,
  author        = {{Maurin}, D. and {Melot}, F. and {Taillet}, R.},
  title         = {{A database of charged cosmic rays}},
  journal       = "Astron. Astrophys.",
  year          = {2014},
  volume        = {569},
  pages         = {A32},
  month         = sep,
  url        = {https://ui.adsabs.harvard.edu/abs/2014A%26A...569A..32M},
  archiveprefix = {arXiv},
  doi           = {10.1051/0004-6361/201321344},
  eid           = {A32},
  eprint        = {1302.5525},
  primaryclass  = {astro-ph.HE},
}

@ARTICLE{MaurinEtAl2020,
       author = {{Maurin}, David and {Dembinski}, Hans Peter and {Gonzalez}, Javier and
         {Mari{\textcommabelow s}}, Ioana Codrina and {Melot}, Fr{\'e}d{\'e}ric},
        title = "{Cosmic-Ray Database Update: Ultra-High Energy, Ultra-Heavy, and Antinuclei Cosmic-Ray Data (CRDB v4.0)}",
      journal = {Universe},
         year = 2020,
        month = jul,
       volume = {6},
       number = {8},
        pages = {102},
          doi = {10.3390/universe6080102},
archivePrefix = {arXiv},
       eprint = {2005.14663},
 primaryClass = {astro-ph.HE},
       adsurl = {https://ui.adsabs.harvard.edu/abs/2020Univ....6..102M}
}

@ARTICLE{MaurinEtAl2023,
       author = {{Maurin}, David and {Ahlers}, Markus and {Dembinski}, Hans and {Haungs}, Andreas and {Mangeard}, Pierre-Simon and {Melot}, Fr{\'e}d{\'e}ric and {Mertsch}, Philipp and {Wochele}, Doris and {Wochele}, J{\"u}rgen},
        title = "{A cosmic-ray database update: CRDB v4.1}",
      journal = {European Physical Journal C},
         year = 2023,
        month = oct,
       volume = {83},
       number = {10},
          eid = {971},
        pages = {971},
          doi = {10.1140/epjc/s10052-023-12092-8},
archivePrefix = {arXiv},
       eprint = {2306.08901},
 primaryClass = {astro-ph.HE},
       adsurl = {https://ui.adsabs.harvard.edu/abs/2023EPJC...83..971M}
}

@ARTICLE{1994ApJ...431..705S,
   author = {{Seo}, E.~S. and {Ptuskin}, V.~S.},
    title = "{Stochastic reacceleration of cosmic rays in the interstellar medium}",
  journal = {\apj},
     year = 1994,
    month = aug,
   volume = 431,
    pages = {705-714},
      doi = {10.1086/174520},
   adsurl = {http://adsabs.harvard.edu/abs/1994ApJ...431..705S}
}

@article{DeRomeri:2025dwm,
    author = "De Romeri, Valentina and Donato, Fiorenza and Maurin, David and Stefanuto, Lorenzo and Tolino, Agnese",
    title = "{Antinuclei from primordial black holes}",
    eprint = "2505.04692",
    archivePrefix = "arXiv",
    primaryClass = "hep-ph",
    reportNumber = "CERN-TH-2025-096",
    doi = "10.1103/r6w9-vr6b",
    journal = "Phys. Rev. D",
    volume = "112",
    number = "2",
    pages = "023003",
    year = "2025"
}

@article{Donato:2001eq,
    author = "Donato, Fiorenza and Maurin, David and Taillet, Richard",
    title = "{Beta-radioactive cosmic rays in a diffusion model: test for a local bubble?}",
    eprint = "astro-ph/0108079",
    archivePrefix = "arXiv",
    reportNumber = "LAPTH-862",
    doi = "10.1051/0004-6361:20011447",
    journal = "Astron. Astrophys.",
    volume = "381",
    pages = "539--559",
    year = "2002"
}

@article{Maurin:2014bva,
    author = "Maurin, D. and Cheminet, A. and Derome, L. and Ghelfi, A. and Hubert, G",
    title = "{Neutron monitors and muon detectors for solar modulation studies: Interstellar flux, yield function, and assessment of critical parameters in count rate calculations}",
    eprint = "1403.1612",
    archivePrefix = "arXiv",
    primaryClass = "astro-ph.EP",
    doi = "10.1016/j.asr.2014.06.021",
    journal = "Adv. Space Res.",
    volume = "55",
    pages = "363--389",
    year = "2015"
}

@article{John:2025jln,
    author = "John, Isabelle and Cuoco, Alessandro",
    title = "{Addressing the impact of solar modulation systematic uncertainties on cosmic-ray propagation models}",
    doi = "10.22323/1.501.1304",
    journal = "PoS",
    volume = "ICRC2025",
    pages = "1304",
    year = "2025"
}

@article{Kuhlen:2019hqb,
    author = "Kuhlen, Marco and Mertsch, Philipp",
    title = "{Time-dependent AMS-02 electron-positron fluxes in an extended force-field model}",
    eprint = "1909.01154",
    archivePrefix = "arXiv",
    primaryClass = "astro-ph.HE",
    reportNumber = "TTK-19-34",
    doi = "10.1103/PhysRevLett.123.251104",
    journal = "Phys. Rev. Lett.",
    volume = "123",
    number = "25",
    pages = "251104",
    year = "2019"
}

@article{Long:2024nty,
    author = "Long, Wei-Cheng and Wu, Juan",
    title = "{Probing solar modulation analytic models with cosmic ray periodic spectra}",
    eprint = "2403.20038",
    archivePrefix = "arXiv",
    primaryClass = "astro-ph.SR",
    doi = "10.1103/PhysRevD.109.083009",
    journal = "Phys. Rev. D",
    volume = "109",
    number = "8",
    pages = "083009",
    year = "2024"
}

@article{Duan:2025ead,
    author = "Duan, Kai-Kai and Wang, Xiao and Li, Wen-Hao and Xu, Zhi-Hui and Tsai, Yue-Lin Sming and Fan, Yi-Zhong",
    title = "{Scrutinizing the impact of the solar modulation on AMS-02 antiproton excess}",
    eprint = "2506.13352",
    archivePrefix = "arXiv",
    primaryClass = "astro-ph.CO",
    doi = "10.1088/1475-7516/2025/10/049",
    journal = "JCAP",
    volume = "10",
    pages = "049",
    year = "2025"
}

@article{Aslam:2023gjv,
    author = "Aslam, O. P. M. and Potgieter, M. S. and Luo, Xi and Ngobeni, M. D.",
    title = "{Modulation of Cosmic-Ray Antiprotons in the Heliosphere: Simulations for a Solar Cycle}",
    eprint = "2303.13268",
    archivePrefix = "arXiv",
    primaryClass = "astro-ph.SR",
    doi = "10.3847/1538-4357/ace31e",
    journal = "Astrophys. J.",
    volume = "953",
    number = "1",
    pages = "101",
    year = "2023"
}

@article{Tomassetti:2025nna,
    author = "Tomassetti, Nicola and Bertucci, Bruna and Fiandrini, Emanuele and Khiali, Behrouz",
    title = "{Propagation Times and Energy Losses of Cosmic Protons and Antiprotons in Interplanetary Space}",
    eprint = "2503.14025",
    archivePrefix = "arXiv",
    primaryClass = "astro-ph.HE",
    doi = "10.3390/galaxies13020023",
    journal = "Galaxies",
    volume = "13",
    number = "2",
    pages = "23",
    year = "2025"
}

@article{Cowan_2011,
   title={Asymptotic formulae for likelihood-based tests of new physics},
   volume={71},
   ISSN={1434-6052},
   url={http://dx.doi.org/10.1140/epjc/s10052-011-1554-0},
   DOI={10.1140/epjc/s10052-011-1554-0},
   number={2},
   journal={The European Physical Journal C},
   publisher={Springer Science and Business Media LLC},
   author={Cowan, Glen and Cranmer, Kyle and Gross, Eilam and Vitells, Ofer},
   year={2011},
   month=feb }

@article{Donato:2008jk,
    author = "Donato, F. and Maurin, D. and Brun, P. and Delahaye, T. and Salati, P.",
    title = "{Constraints on WIMP Dark Matter from the High Energy PAMELA $\bar{p}/p$ data}",
    eprint = "0810.5292",
    archivePrefix = "arXiv",
    primaryClass = "astro-ph",
    reportNumber = "DFTT-28-2008, LAPTH-1273-08, IRFU-08-190",
    doi = "10.1103/PhysRevLett.102.071301",
    journal = "Phys. Rev. Lett.",
    volume = "102",
    pages = "071301",
    year = "2009"
}

@article{PhysRevLett.134.051002,
  title = {Antiprotons and Elementary Particles over a Solar Cycle: Results from the Alpha Magnetic Spectrometer},
  author = {Aguilar, M. and others},
  collaboration = {AMS Collaboration},
  journal = {Phys. Rev. Lett.},
  volume = {134},
  issue = {5},
  pages = {051002},
  numpages = {12},
  year = {2025},
  month = {Feb},
  publisher = {American Physical Society},
  doi = {10.1103/PhysRevLett.134.051002},
  url = {https://link.aps.org/doi/10.1103/PhysRevLett.134.051002}
}

@article{G_nolini_2021,
   title={New minimal, median, and maximal propagation models for dark matter searches with Galactic cosmic rays},
   volume={104},
   ISSN={2470-0029},
   url={http://dx.doi.org/10.1103/PhysRevD.104.083005},
   DOI={10.1103/physrevd.104.083005},
   number={8},
   journal={Physical Review D},
   publisher={American Physical Society (APS)},
   author={Génolini, Yoann and Boudaud, Mathieu and Cirelli, Marco and Derome, Laurent and Lavalle, Julien and Maurin, David and Salati, Pierre and Weinrich, Nathanael},
   year={2021},
   month=oct }

@misc{dimauro2026enhancedcosmicrayantinucleifluxes,
      title={Enhanced Cosmic-Ray Antinuclei Fluxes with Dark Matter Annihilation into SUEPs}, 
      author={Mattia Di Mauro and Caleb Gemmell and Austin Batz and David Curtin and Fiorenza Donato and Nicolao Fornengo and Graham D. Kribs},
      year={2026},
      eprint={2602.15132},
      archivePrefix={arXiv},
      primaryClass={hep-ph},
      url={https://arxiv.org/abs/2602.15132}, 
}

@Article{instruments6020019,
AUTHOR = {Adriani, Oscar and others},
TITLE = {Design of an Antimatter Large Acceptance Detector In Orbit (ALADInO)},
JOURNAL = {Instruments},
VOLUME = {6},
YEAR = {2022},
NUMBER = {2},
ARTICLE-NUMBER = {19},
URL = {https://www.mdpi.com/2410-390X/6/2/19},
ISSN = {2410-390X},
DOI = {10.3390/instruments6020019}
}

@article{SAFFOLD2021102580,
title = {Cosmic antihelium-3 nuclei sensitivity of the GAPS experiment},
journal = {Astroparticle Physics},
volume = {130},
pages = {102580},
year = {2021},
issn = {0927-6505},
doi = {https://doi.org/10.1016/j.astropartphys.2021.102580},
url = {https://www.sciencedirect.com/science/article/pii/S0927650521000244},
author = {N. Saffold and others},
collaboration = {GAPS},
}

@Misc{berlillium2026,
  author       = {J. O. Peleteiro},
  howpublished = {ECRS 2026},
  title        = {Properties of Cosmic Beryllium Isotopes Measured by the Alpha Magnetic Spectrometer},
  year         = {2026},
  url          = {https://indico.cern.ch/event/1536846/contributions/7101744/attachments/3307357/5917831/ECRS26_AMS_BerylliumIsotopes.pdf},
}

\end{document}